\documentclass[preprint]{aastex631}

\newcommand{\fe}{Fe\,\textsc{xxi}\ }

\newcommand{\rfont}{
}
\DeclareTextFontCommand{\revise}{\rfont}
\DeclareTextFontCommand{\edit}{\rfont}

\shortauthors{Ashfield et al.}

\graphicspath{{./}{figures/}}

\begin{document}

\title{Non-thermal Observations of a Flare Loop-top using IRIS Fe XXI: Implications for Turbulence and Electron Acceleration}

\accepted{July 16th, 2024}
\submitjournal{ApJ}

\author[0000-0002-6368-939X]{William Ashfield IV}
\affiliation{Bay Area Environmental Research Institute, NASA Research Park, Moffett Field, CA 94035-0001, USA}
\affiliation{Lockheed Martin Solar and Astrophysics Laboratory, Building 203, 3251 Hanover Street, Palo Alto, CA 94304, USA}

\author[0000-0002-4980-7126]{Vanessa Polito}
\affiliation{Lockheed Martin Solar and Astrophysics Laboratory, Building 203, 3251 Hanover Street, Palo Alto, CA 94304, USA}
\affiliation{Oregon State University, Department of Physics, 301 Weniger Hall, Corvallis, 97331, OR, USA }

\author[0000-0003-2872-2614]{Sijie Yu}
\affiliation{Center for Solar-Terrestrial Research, New Jersey Institute of Technology, 323 M L King Jr Boulevard, Newark, NJ 07102-1982, USA}

\author[0000-0001-5592-8023]{Hannah Collier}
\affiliation{University of Applied Sciences and Arts Northwestern Switzerland, Bahnhofstrasse 6, 5210 Windisch, Switzerland}
\affiliation{ETH Zürich, Rämistrasse 101, 8092 Zürich Switzerland}

\author[0000-0002-6835-2390]{Laura A. Hayes}
\affiliation{European Space Agency, ESTEC, Keplerlaan 1 - 2201 AZ, Noordwijk, The Netherlands}

\begin{abstract}
The excess broadening of high-temperature spectral lines, long observed near the tops of flare arcades, is widely considered to result from magnetohydrodynamic (MHD) turbulence. 
According to different theories, plasma turbulence is also believed to be a candidate mechanism for particle acceleration during solar flares. However, the degree to which this broadening is connected to the acceleration of non-thermal electrons remains largely unexplored outside of recent work, and many observations have been limited by limited spatial resolution and cadence. Using the \textsl{Interface Region Imaging Spectrometer} (IRIS), we present spatially resolved observations of loop-top broadenings using hot ($\approx11$\,MK) Fe\,\textsc{xxi} 1354.1\,\AA\ line emission at $\approx9$\,s cadence during the 2022 March 30 X1.3 flare. We find non-thermal velocities upwards of 65\,km\,s\(^{-1}\) that decay linearly with time, indicating the presence and subsequent dissipation of plasma turbulence. Moreover, the initial Fe\,\textsc{xxi} signal was found to be co-spatial and co-temporal with microwave emission measured by the \textsl{Expanded Owens Valley Solar Array} (EOVSA), placing a population of non-thermal electrons in the same region as the loop-top turbulence. Evidence of electron acceleration at this time is further supported by hard X-ray measurements from the \textsl{Spectrometer/Telescope for Imaging X-rays} (STIX) aboard Solar Orbiter. Using the decay of non-thermal broadenings as a proxy for turbulent dissipation, we found the rate of energy dissipation to be consistent with the power of non-thermal electrons deposited into the chromosphere, suggesting a possible connection between turbulence and electron acceleration. 

\end{abstract}

\section{Introduction} \label{sec:intro}

According to the standard model of solar flares, magnetic free energy stored in the corona is released through the reconfiguration of highly stressed field lines via magnetic reconnection. This energy is ultimately converted to other forms and transported throughout the solar atmosphere \citep{syrovatskii1972,craig1976,kopp1976}, driving the phenomena observed with these events. One such transport mechanism at the heart of the standard model is the acceleration of non-thermal electrons from the reconnection site to the lower atmosphere, which in turn allows for the heating and ablation of chromospheric plasma upwards into newly reconnected flare loops during the evaporation process. While indirect evidence for this process is observed as the increases in soft X-ray (SXR) and extrema ultraviolet (EUV) emission of flare loops \cite{neupert1968}, direct evidence of non-thermal electron deposition manifests in the form of hard X-ray emission (HXR) at the location of flare footpoints in the chromosphere \cite{brown1971}.

Although there is much evidence for non-thermal electron transport from the corona to the lower atmosphere \citep{aschwanden1996,holman2011}, the mechanism behind particle acceleration is still in question. Popular theories regarding this process are linked to the dynamics of the reconnection process, such as the formation of magnetic islands and plasmoids \citep{drake2006}, or termination shocks following the impact of reconnection outflows at the base of the current sheet \citep{forbes1986,chen2015}. These outflows, especially in the regions stemming from the flare arcade loop-top (LT) and above, are also likely to generate magnetohydrodynamic (MHD) turbulence given the large Reynolds number of coronal plasma, as has been seen in several numerical studies \citep{ruan2018,ye2020,shen2023_2},\revise{in addition to observation \citep{mckenzie2013,cheng2018}}. The cascade of energy to smaller scales in MHD turbulence can also energize electrons on the kinetic scale, providing a stochastic transfer of energy to electrons that would result in their acceleration \citep{moore1993,goldreich1995,miller1997,petrosian2012}. Recent microwave (MW) spectral imaging data taken with the Expanded Owens Valley Solar Array \citep[EOVSA;][]{gary2018} has brought about a paradigm shift in high-energy flare physics, allowing for spatially and temporally resolved measurements of non-thermal electrons and magnetic fields through gyrosynchrotron emission of accelerated electrons in the flaring corona. Although recent work using EOVSA has provided evidence for the acceleration site location near the post-flare LTs \citep{chen2020,yu2020,fleishman2022}, additional diagnostics are necessary to constrain the acceleration mechanisms involved.

Given the high-temperature nature of flare plasma, spectroscopic observations of ultraviolet (UV) emission lines — sensitive to such high-temperature plasma — have shed much light on the properties of plasma dynamics during flares, such as mass flows and electron densities (see reviews by \cite{fletcher2011} and \cite{delzanna2018}). An observable of UV spectra that continues to remain of particular interest to the flare community is the degree to which measured line profiles broaden beyond their respective thermal widths. Due to the fluctuations in the plasma velocity field, either from large-scale bulk plasma flows or microscopic ion perturbations, the presence of enhanced line broadening is widely believed to be a signature of MHD turbulence \citep[e.g.][]{milligan2011,polito2019}.

While the signatures of turbulence in solar flares remain a topic of active research, investigations have started exploring the evolution of non-thermal broadenings in the context of turbulent dissipation. Starting with \cite{kontar2017}, measurements in non-thermal broadening were found to decay over time, which was used to infer a rate of turbulent kinetic energy dissipation. This energy was found to be consistent with the power of non-thermal elections as measured by RHESSI, indicating a possible connection between MHD turbulence and the coronal electron acceleration site. However, the study was limited to values of non-thermal line width averaged over the total emission region using the Extreme-ultraviolet Imaging Spectrometer (EIS). A follow-up study by \cite{stores2021} mapped the spatial distribution and evolution of non-thermal velocities of the same event at a resolution of 4\arcsec, confirming that turbulence as inferred from such velocities is not necessarily localized, but can vary along the LT and loop-leg regions of a post flare arcade. The most recent work by \cite{shen2023_2} used observations from the Interface Region Imaging Spectrometer \citep[IRIS;][]{depontieu2014} to study the evolution of non-thermal broadenings in the IRIS \fe 1354.08\,\AA\  line and their subsequent decay in the context of MHD turbulence. Using a three-dimensional MHD simulation, they found turbulent bulk plasma flows in the current sheet and flare LT regions were responsible for the non-thermal broadening of the \fe emission line. Despite this progress, the decay of non-thermal broadenings and their subsequent connection to the acceleration site of non-thermal electrons remains largely unexplored.

The present investigation undertakes a multi-instrument approach to study the evolution of non-thermal signatures associated with the March 30th, 2022 X1.3 flare. 
Using observations with high spatial and spectral resolution measured by IRIS, we find instances where the \fe spectral line displays relatively large non-thermal widths at the flare LT, with non-thermal velocities reaching upwards of 65\,km\,s$^{-1}$.  By tracking the evolution of these broadenings, we find two periods of time where the non-thermal broadenings decrease linearly with time, an evolution characteristic of MHD turbulent decay. Moreover, the initial \fe signal is co-spatial and co-temporal with gyrosynchrotron microwave emission as measured with EOVSA. When combined with HXR measurements taken from the Spectrometer/Telescope for Imaging X-rays (STIX) aboard Solar Orbiter (SolO), we find this activity indicates a population of non-thermal electrons along the loop in conjunction with electron deposition in the chromosphere. As these non-thermal signatures exist in unison, we aim to explore their possible relationships.

The paper is organized as follows. An overview of the flare and the observational data used in this study is presented in Section \ref{sec:obs}, followed by an analysis of the \fe line profiles and their implications for the flare LT dynamics presented in Section \ref{sec:fe21}. Signatures of electron acceleration using EOVSA and STIX in relation to the initial \fe emission are analyzed in Section \ref{sec:accel}. We discuss the implications of the decay of non-thermal broadenings in terms of kinetic energy dissipation and its connection to the acceleration site in Section \ref{sec:turb}. A brief summary of the results and conclusions is given in Section \ref{sec:disc}.

\section{Flare Overview} \label{sec:obs}

\begin{figure}[!t]
    \centering
    \includegraphics[width = 0.66\textwidth]{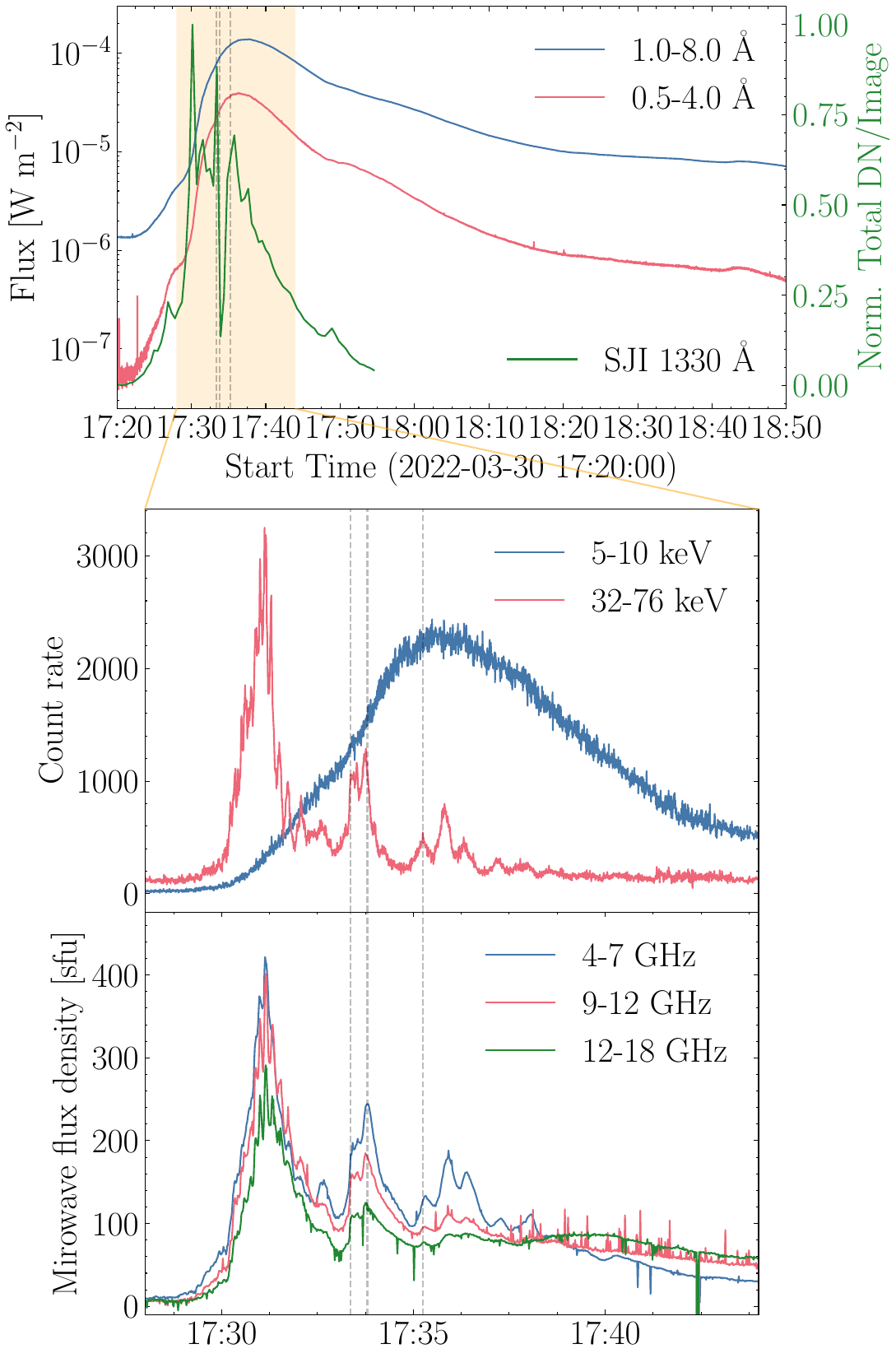}
    \caption{Light curve overview of the 2022 March 30 X1.3-class flare. GOES SXR fluxes (blue and red lines) and normalized SJI 1330\,\AA \, counts (green) in the top panel illustrate the evolution of the flare captured by IRIS. The two enlarged panels on the bottom show the HXR (red) and SXR (blue) count rate from SolO/STIX, and microwave flux densities in three frequency bands from EOVSA, respectively. The vertical dashed lines indicate the start time of the initial loop-top \fe emission, the peak time of the second phase, and the first pulse in the third phase, respectively, and correspond to the three times in Figure \ref{fig:eovsa}. 
    \label{fig:tmseries} }
\end{figure}

The March 30th 2022 event (SOL2022-03-30T17:23:06) was an X1.3 GOES class eruptive flare in NOAA active region AR 12975, located on the NW quadrant of the solar disk as viewed from Earth. 
From the GOES X-ray Sensor (XRS) 1-8\,\AA\, filter light curve in Figure \ref{fig:tmseries}, the flare began at $\approx$17:20 UT, peaked at 17:37:42\,UT, and lasted roughly 1.5 hours. Throughout its lifetime, the flare was well observed in EUV by Atmospheric Imaging Assembly aboard the Solar Dynamics Observatory \citep[SDO/AIA;][]{lemen2012}, 4-150\,keV X-rays by SolO/STIX \citep{krucker2020}, and 1-18\,GHz MW by EOVSA \citep{gary2018}. 

Beginning in the impulsive phase, STIX observed high-energy signatures in three discrete HXR emission pulsations, or \textsl{phases} \citep{collier2023}. Figure \ref{fig:tmseries} shows the HXR emission count rate over the 32-76\;keV energy bin at a 0.5\,s cadence. 
Given SolO's near-perihelion location at 0.33\,AU, reference times have been adjusted to Earth's reference time at UT. Likewise, full-disk integrated MW dynamic spectra taken with EOVSA show three distinct intensity phases. The bottom panel of Figure 1 plots MW flux in three GHz bins covering the effective broadband range of EOVSA, each correlating with the phases and individual bursts seen in the 32-76\;keV STIX channels. As gyrosynchrotron radiation from coronal electrons dominates EOVSA MW emission, the degree to which the MWs phases correlate with the HXR emission emphasizes its use as a valuable diagnostic for electron acceleration.\revise{The following work focuses on times occurring over the second and third energy phases, which coincide with MW emissions occurring in proximity to the IRIS SG slit.}  A more in-depth investigation of HXR and MW pulsations during the first phase can be found in recent work by \cite{collier2024}.

Observations taken with IRIS \citep{depontieu2014,depontieu2021} captured most of the flare's evolution in the UV, with the observing run ending at 17:54:31 (see the green curve in Figure \ref{fig:tmseries}). Running in a sit-and-stare mode (OBSID 3660259102), the spacecraft captured slit-jaw images (SJI) at a cadence of 28.1 seconds in the 1330\,\AA\,channel. A flare line list spectral readout including the O~\,\textsc{i} window (1351.3 - 1356.46 \AA) had a variable exposure time ranging from 0.8-8.0 seconds and a cadence of 3.9-14.7\,s. The spatial resolution of the SJI and spectrograph (SG) slit was 0.166\arcsec\,per pixel, with a 0.33\arcsec\,wide slit. The FUV1 channel (1331.7–1358.4) was binned spectrally by two onboard, resulting in a resolution of 26 m\AA.

\begin{figure*}
    \centering
    \includegraphics[width = \textwidth]{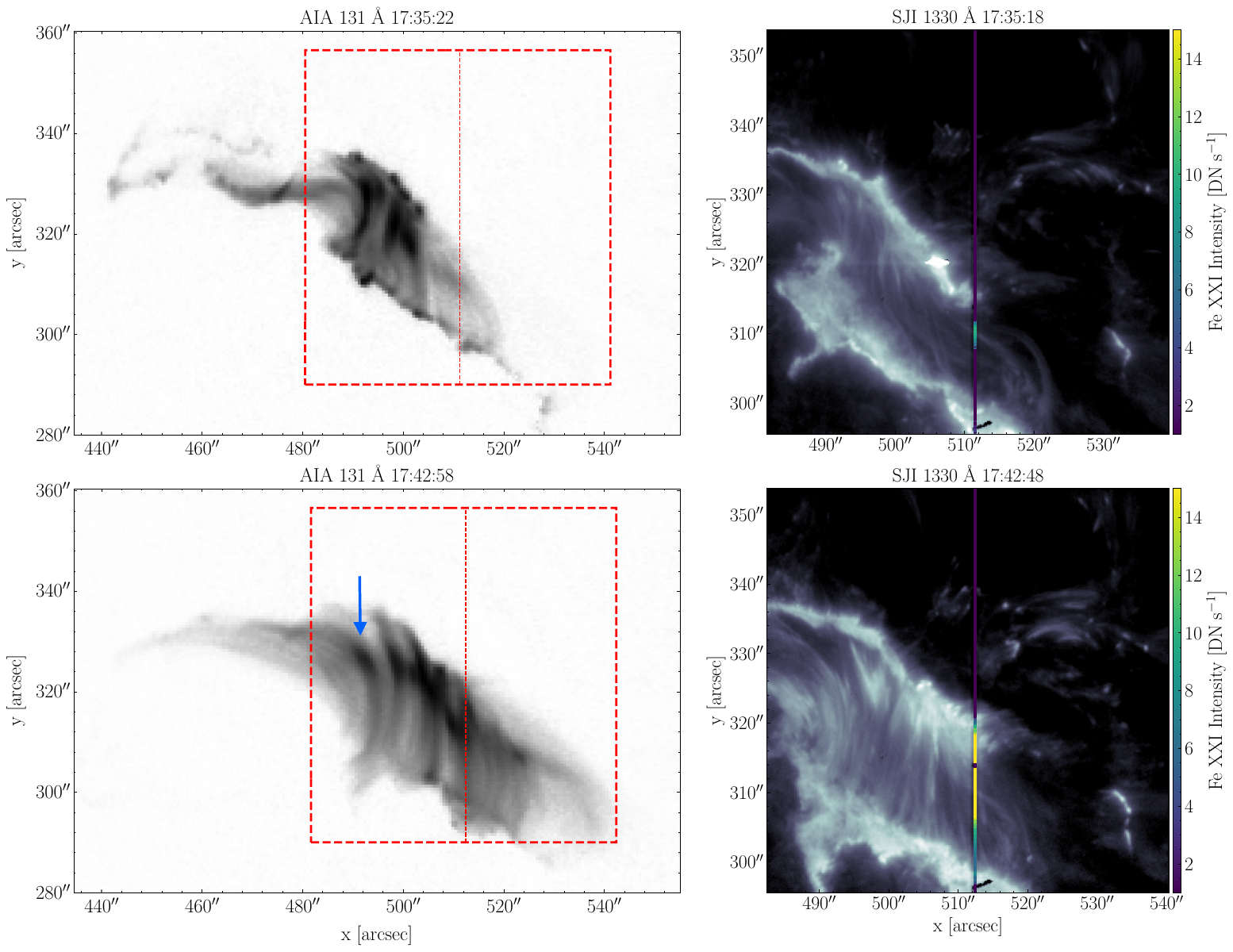}
    \caption{AIA~131\,\AA \, images in inverse logarithmic greyscale (\revise{left}), along with corresponding IRIS SJI 1330\,\AA \, images (\revise{right}), showing the evolution of the flare arcade at two times. The red boxes and dashed lines show the SJI FOV and slit position, respectively.\revise{EUV knots correspond to the brightest (most-black) emission along the LT, with an example highlighted by the blue arrow.} The multi-color line gives the peak \fe intensity measured along the IRIS SG slit.
    \label{fig:aia_sji} }
\end{figure*}

The IRIS field of view (FOV) was favorably positioned over the post-flare loop arcade during its observing run, as shown in the right panels in Figure \ref{fig:aia_sji}. Here, the arcade is given by images of AIA~131\,\AA\ and illustrates the morphology of hot coronal plasma over time. Because the event was a so-called two-ribbon flare, with the parallel ribbons roughly running in the east-west direction, the location of the slit allowed for complete spectral measurements of the southern ribbon and the westward edge of the northern ribbon. Moreover, the slit position aids in the identification of the LT emission, as spectra emanating between the two ribbons could belong to arcade plasma. The following section explores the behavior of \fe spectra in this region. 

\section{Loop-top Fe \textsc{XXI} Emission} \label{sec:fe21}

The IRIS \fe1354.1\,\AA\, emission line, formed at temperatures of $\approx$11.5\,MK, allows for spectrographic observations of the flaring corona. Like many optically thin spectra, \fe is frequently observed to have a Gaussian line profile, given an isothermal Maxwellian distribution of plasma ions \citep[e.g.][]{graham2015,polito2019}. Following an initial verification of the line profile shapes, we perform Gaussian fits to the \fe spectra to infer plasma motion diagnostics stemming from the arcade LT.

Prior to fitting, level 2 raster data were processed starting at 17:24:31\,UT, preceding the start of the impulsive phase, and lasting until the end of the observation run 30 minutes later. The lower IRIS SG fiducial mark, visible as dark bands on SG detector images, overlapped regions of interest in our observations and was excluded by ignoring pixels 142-144 (314.4\arcsec-314.9\arcsec) throughout our analysis to ensure no related intensity artifacts. The SG data was further normalized by exposure time to remove any effects of variable exposure time and cleaned via the de-spiking procedure \texttt{iris\_prep\_despike} in SolarSoft to remove bad pixels \citep{freeland1998}. Lastly, because the \fe signal was relatively weak throughout the observation, an additional spectral binning of two was applied to the SG data to reduce noise for the fitting process, resulting in a 52\,m\AA\, spectral resolution.

Although LT emission is the focus of this work, we aim to fit \fe spectra along the entire SG slit, including emission from the flare ribbons. Previous observations have shown that \fe ribbon emission is blended with other chromospheric and photospheric lines in the O\,\textsc{i} channel, including the predominant the C\,\textsc{i}\,1354.3\,\AA\ line \citep{polito2015,polito2016}. We therefore employed the \texttt{iris\_auto\_fit} multi-Gaussian routine\footnote{Further details can be found at \url{https://www.pyoung.org/quick_guides/iris_auto_fit.html} } in SolarSoft to account for the presence of these lines.
Prior to the routine, weak signals with wavelength-integrated intensity $\pm$0.5\AA\ of the \fe line below a threshold of 50 DN\,\AA\,s$^{-1}$ were ignored from the fit. Likewise, the fit was discarded from the results if the Gaussian amplitude, or peak intensity, was less than twice the measured background intensity at a given time. 

\begin{figure*}
    \centering
    \includegraphics[width = 0.9\textwidth]{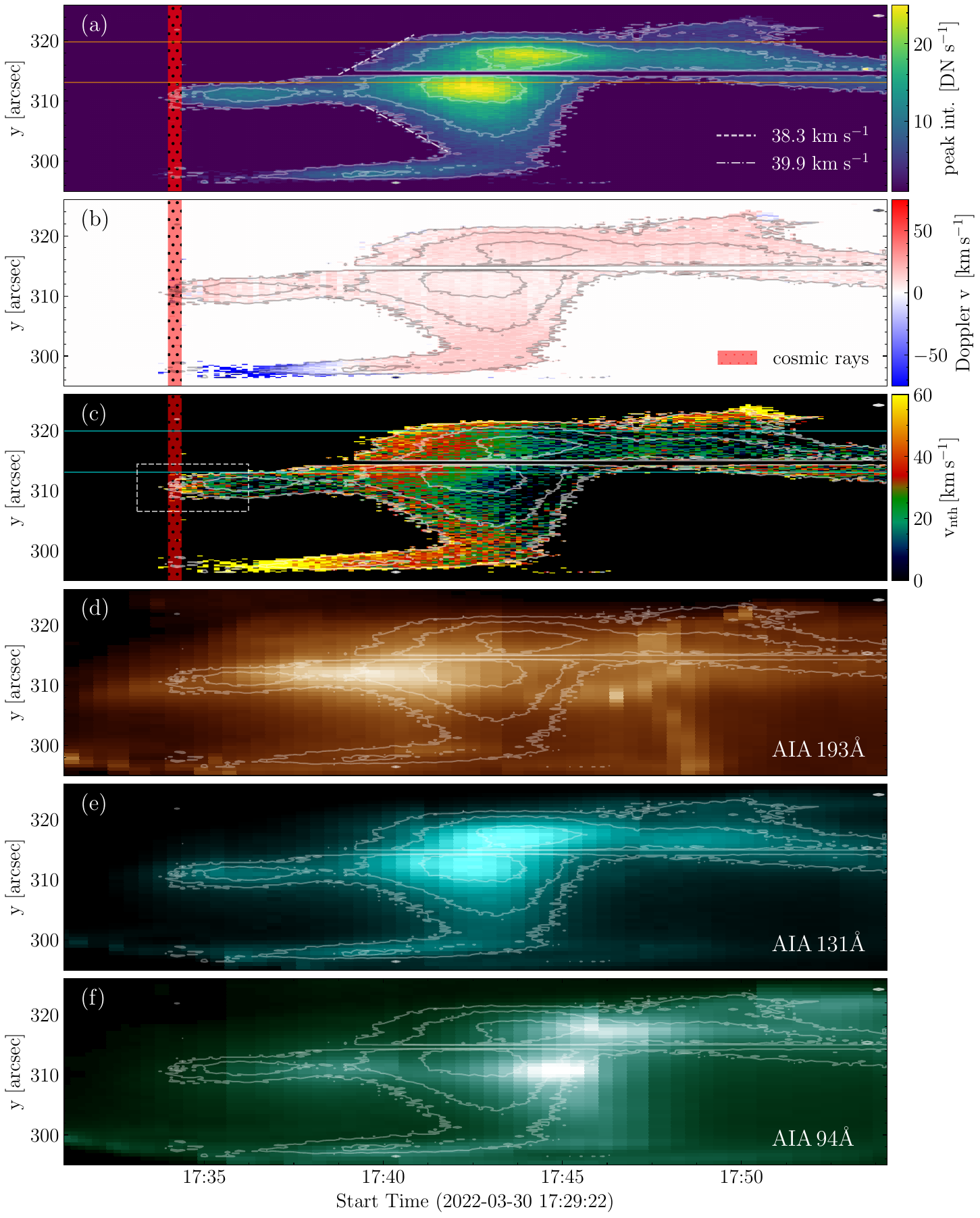}
    \caption{Time-distance stack plots of \fe spectral line parameters and AIA emission in three channels along the IRIS SG slit. The top three panels show intensity (a), Doppler velocity (b), and FWHM line width (c) as inferred from Gaussian fits to \fe. Dashed and dashed-dotted lines give the northward and southward \fe emission propagation speed away from the LT, respectively.\revise{The orange and cyan lines in (a) and (c), respectively, outline the pixels used in Figure \ref{fig:aligned}}.  The white dashed box (c) outlines the time and location of the initial LT \fe signal. Times of high CCD noise due to increased cosmic ray exposure are masked in red. Overlaid contours outline the \fe intensity at [4,9,18] DN\,s$^{-1}$.
    \label{fig:stack} }
\end{figure*}

Figure \ref{fig:stack} shows time-distance stack plots of the resulting plasma diagnostics inferred using \fe spectral line parameters along the IRIS SG slit. The top three panels (a-c) show peak intensity, Doppler velocity, and non-thermal line width, respectively. Doppler velocities were calculated using the differences between Gaussian centroid positions and the rest wavelength, taken here to be 1354.08\,\AA\ following several studies of this line \citep{sandlin1986,polito2015}. Non-thermal widths were calculated using the FWHM of the Gaussian fits, where FWHM = 2$\sqrt{2\ln2}$\,$\sigma$ for Gaussian line width $\sigma$. By decomposing the line width into thermal, non-thermal, and instrumental components, the FWHM can be expressed as
\begin{equation}
    \mathrm{FWHM} = \sqrt{
    4\ln2\Bigg(\frac{\lambda_0}{c}\Bigg)^2 \Big(v_\mathrm{th}^2 + v_\mathrm{nth}^2   \Big) + \mathrm{FWHM}_\mathrm{IRIS}^2
    },
\end{equation}
where $v_\mathrm{nth}$ is the non-thermal broadening expressed as a velocity. The thermal velocity $v_\mathrm{th}$ is given by the Doppler broadening of the line, $\sqrt{2k_b T_\mathrm{Fe}/m_i}$, for ion mass $m_i$=56 and formation temperature $T_\mathrm{Fe}$. Here, we take Log$_{10}\,T_\mathrm{Fe}$[K] = 7.06, the temperature at the contribution function maximum calculated using Chianti v10 \citep{delzanna2021}, which gives a thermal width of 0.439\,\AA\,= 58.4\,km\,s$^{-1}$. Broadening due to instrumental effects is given by $\mathrm{FWHM}_\mathrm{IRIS}$ = 26\,m\AA. The flight time of IRIS through the Southern Atlantic Anomaly (SAA) during the observation run, resulting in increased cosmic ray exposure and SG noise, is highlighted in red.

\fe emission along the slit can be considered to evolve from two separate regions: the LT and the ribbons. The southern ribbon, starting at $y\approx295$\arcsec\,, is outlined by the strong Doppler velocity blueshifts in Figure \ref{fig:stack}(b). These large upflows, reaching speeds upwards of 130\,km\,s$^{-1}$  provide clear evidence of chromospheric evaporation. While the northern ribbon is not as clearly defined here, chromospheric lines in the O\,\textsc{i} detector images show its location starting at $y\approx317$\arcsec\, and spreading northward over time (Figure \ref{fig:db_frame}). A more detailed ribbon analysis of this flare can be found in recent work by \cite{xu2023} and \cite{wang2023}.

We identify the LT emission by the distinct increase in \fe intensity at $y\approx311$\arcsec, which brightens almost in tandem with the ribbon emission. As mentioned above, emission in the region between the ribbons in a two-ribbon flare could belong to the post-flare arcade loops. This idea can be more readily seen by the location of the \fe intensity along the slit illustrated in Figure \ref{fig:aia_sji}, where the initial signal of the emission does not emanate from either ribbon. Moreover, the LT \fe profiles themselves support a coronal source by lacking the blend photospheric lines that are typical of ribbon emission. Although this notion constrains the emission to the coronal plasma, a more rigorous definition of LT emission is provided by comparing \fe intensity to AIA~131\,\AA\,emission. As seen in\revise{Figure} \ref{fig:aia_sji}, the morphology of AIA~131\,\AA\, is characterized by bright, localized LT structures, commonly referred to as EUV knots. These knots have long been observed along the arcades of post-flare loops \citep{widing1974,acton1992}, and while their formation mechanism remains unclear \citep[e.g.][]{patsourakos2004,reeves2007}, they are widely considered to exist at the apex of such loops.

To study the connection between knots and \fe emission in this study, slices along AIA images in three wavelengths (94~\AA,131~\AA, and 193~\AA) were taken over time, with the artificial slit interpolated in space to match the IRIS slit position at AIA observation times. We note that all AIA images used in this work were first prepped using routines found in the aiapy Python package \citep{barnes2020}, filtered to select only non-saturated data, and deconvolved using the instrumental PSF. The resulting time-distance stack plots are shown in Figure \ref{fig:stack}(d)-(f), with the loop-top knots corresponding to the regions of peak intensity. Notably, the knots in each channel brighten sequentially, hinting at a cooling plasma provided the characteristic formation temperatures of the three AIA channels --- Log$_{10}\,T$ [K]  = 7.3,7.0,6.8 for AIA 193~\AA, 131~\AA, and 94~\AA, respectively --- during flares \citep{odwyer2010}. Focusing on the 131~\AA\,channel in Figure \ref{fig:stack}(e), we found the evolution to neatly match that of the \fe peak intensity (grey contours). While this result is not surprising given that AIA~131\,\AA\, is dominated by \fe~128.75~\AA\ during flares \citep{odwyer2010}, the similarity between the two emission sources along the SG slit shows that the IRIS \fe emission is also capturing the evolution of the AIA LT knot, thereby acting as a spectroscopic proxy by providing plasma diagnostics for the AIA~131\,\AA\ channel.

\begin{figure}
    \centering
    \includegraphics[width = 0.7\columnwidth]{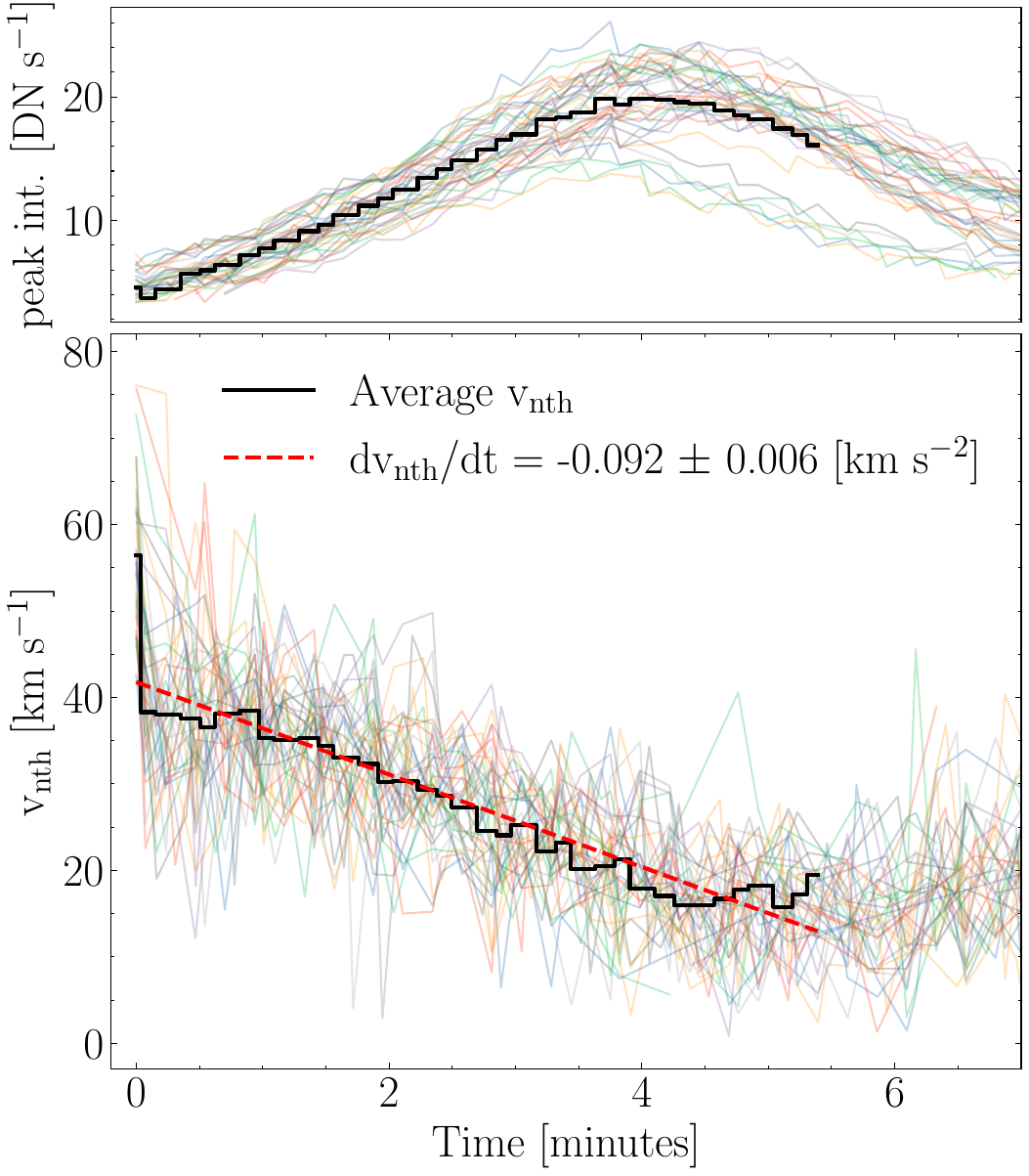}
    \caption{Time series of peak intensities (top) and non-thermal velocities (bottom) from SG slit pixels corresponding to the region [312.9\arcsec-319.9\arcsec] in Figure \ref{fig:stack} (a) and (c), respectively. Pixels are aligned in time according to their respective activation time. Black curves signify the average values in each plot. The bottom panel shows the best linear fit (red-dashed) when $v_\mathrm{nth}$ is decreasing. 
    \label{fig:aligned} }
\end{figure}

Several aspects of the LT dynamics inferred from \fe are worth noting. First, following a period of stationary emission, the signal begins to spread outward along the slit at nearly constant velocities, as illustrated in Figure \ref{fig:stack}(a). This movement implies the top-down motion of hot flare plasma from the LT towards the ribbons, and in particular, the evaporation signatures from the southern ribbon appear to connect with the southward moving LT front much lower down at $y\approx302$\arcsec. 
Second, the entire LT structure is weakly redshifted. Velocities between 304.1\arcsec - 319.9\arcsec range from 0.1-15.5\,km\,s$^{-1}$, with average value of 7.8\,km\,s$^{-1}$. Lastly, the bright LT knots are preceded by a period of non-thermal line broadening, as shown in Figure \ref{fig:stack}(c). In particular, there exists an apparent gradient in $v_\mathrm{nth}$ over time, beginning as the LT emission starts to spread along the slit. This gradient is more clearly seen in Figure \ref{fig:aligned}, where $v_\mathrm{nth}$ and peak intensities from each pixel in the region [312.9\arcsec-319.9\arcsec] are aligned according to their respective activation time (e.g. when the intensity first appears). Averaging over these pixels, we find the gradient is well described by a linear fit, with a decay rate of -0.092$\pm$0.006\,km\,s$^{-2}$, and has an inverse relationship with the increase in peak intensity observed in the same period. This decay in $v_\mathrm{nth}$ also complements earlier observations made by \citep{young2015} regarding a two-ribbon flare, where they did not detect any significant line broadening along the loops at the points where the IRIS raster and AIA~131\,\AA\, knots overlapped. Comparing the widths in their Figure 12 to the non-thermal velocities in our Figure \ref{fig:stack}(c), we see that the broadening at the peak time and location of the LT knots are comparable. Moreover, they found the non-thermal velocities of the \fe line to decrease over time, falling from 42.8 to 26.3\,km\,s$^{-1}$ over the course of 6 minutes by taking the median value of each 8-step raster at a cadence of 75\,s. By using a sit-and-stare observation, this work allows for a more comprehensive analysis of this non-thermal decay.

Large values of non-thermal velocities are also seen as the LT \fe emission becomes visible. The initial signal also appears to precede the formation of the EUV knot, which first manifests in the AIA 193\,\AA\, channel (Figure \ref{fig:stack}(d)) at $\approx$17:37:00. As this signal is separate from the knot evolution described above, a closer look is warranted to understand the nature of the LT emission. 

\begin{figure}[t]
    \centering
    \includegraphics[width = \columnwidth]{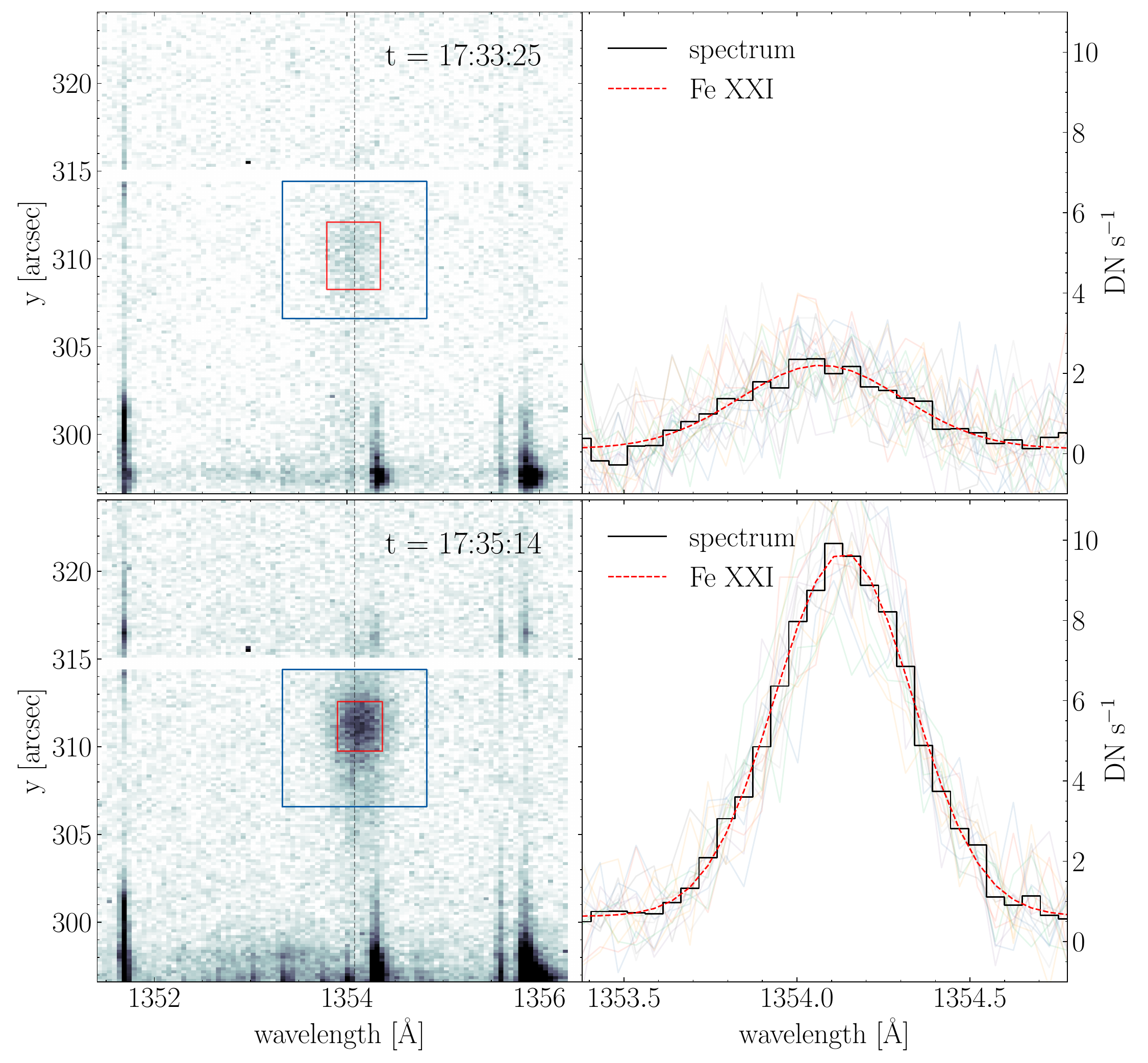}
    \caption{Examples of the dynamic binning routine. IRIS detector images of the O\,\textsc{i} channel in inverse greyscale are shown on the left. Grey dashed lines give the rest wavelength of \fe 1354.08\,\AA, and blue boxes outline the applied region. 
    The red box outlines the selected signal, with the height and width denoting the spatially binned pixels and the resulting FWHM of the Gaussian fit to the binned spectra, respectively. The binned spectra (black) and Gaussian fits (red dashed) are shown on the right, overtop the selected individual spectra. 
    \label{fig:db_frame} }
\end{figure}

\begin{figure}
    \centering
    \includegraphics[width = 0.8\columnwidth]{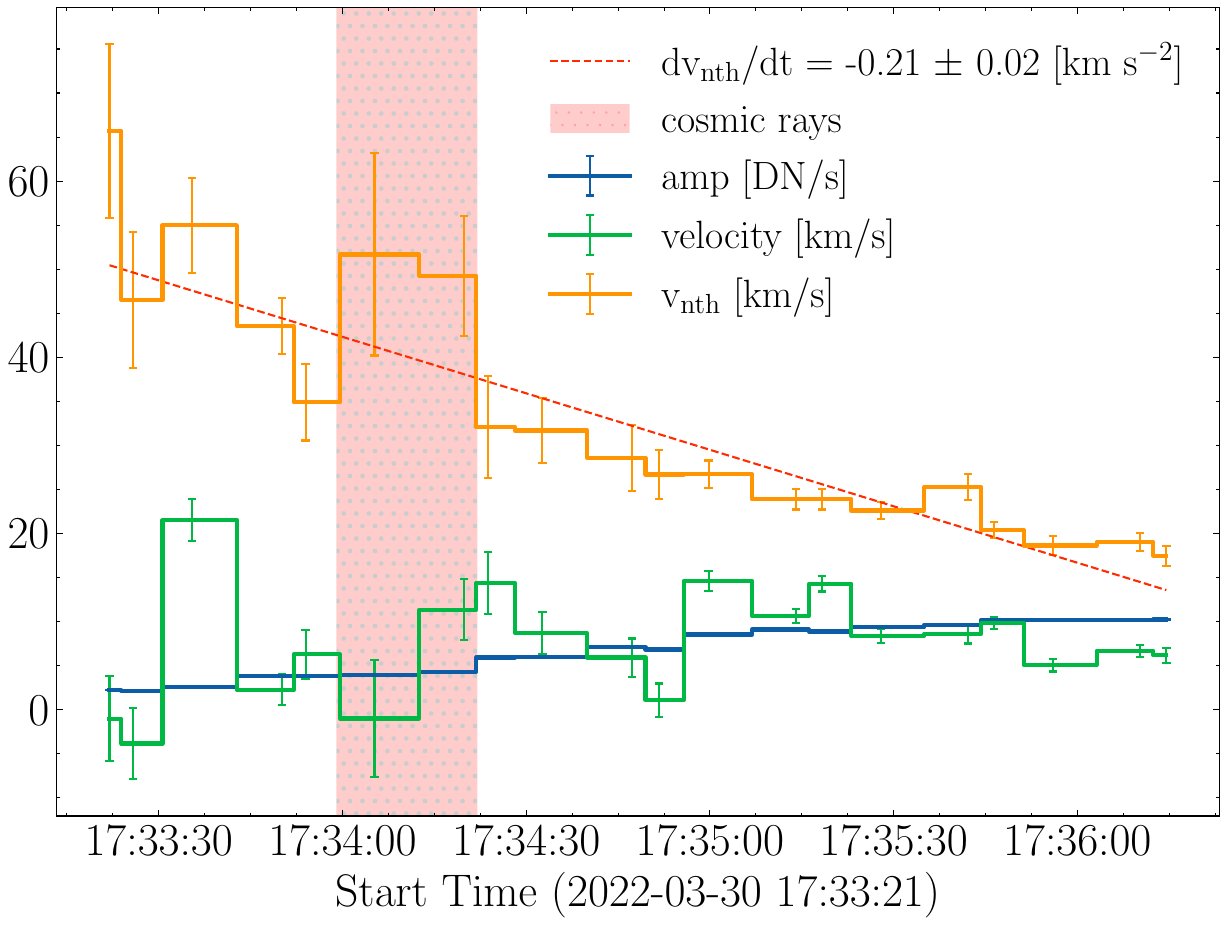}
    \caption{Time series of \fe spectral line parameters from the dynamic binning routine over the initial signal. Lines show peak intensity (blue), Doppler velocity (green), and non-thermal velocity (orange). The red dashed line shows the linear best fit to the non-thermal velocity. Times of increased cosmic ray exposure are masked in red.
    \label{fig:fexxi_tmseries} }
\end{figure}

\subsection{Initial Signal} \label{sec:is}

Although the LT \fe emission in Figure \ref{fig:stack} starts at around 17:34:30, closer inspection of O\,\textsc{i} detector images shows the \fe signal appearing at 17:33:25. To fix this discrepancy, due to the signal threshold selection criteria in the multi-Gaussian fitting routine, we performed a second fitting routine that automatically binned the spectra in space according to the signal-to-noise ratio (SNR) of the \fe line. Two examples\revise{of} this routine, henceforth the dynamic binning routine, can be found in Figure \ref{fig:db_frame}. The routine is applied to a region in detector image space corresponding to the LT emission along the slit (306.6\arcsec-314.2\arcsec) and spanning $\pm0.75$ of \fe line center (blue box). Within that region, pixels with signals greater than 0.5$\sigma$ above the region are flagged to identify the \fe emission (height of the red box) and are subsequently binned to create a single spectrum (black curve) that is then fit with a Gaussian (red dashed curve). The FWHM of the Gaussian fit determines the width of the red box. We note that a single Gaussian was sufficient here, as no chromospheric line blends were detected in the \fe spectra. By automatically binning the spectra in this way, we increase the SNR without including superfluous signals and mitigate the loss of spatial resolution.

The dynamic binning routine was applied on the SG data at each time during the initial signal from 17:33:25-17:36:14. Figure \ref{fig:fexxi_tmseries} shows the time series of the \fe spectral line fit parameters. The LT emission begins with a relatively high non-thermal velocity at 65.7\,km\,s$^{-1}$, which decreases steadily to 17.4 \,km\,s$^{-1}$ in less than three minutes. Like the line width decay preceding the EUV knots (Figure \ref{fig:aia_sji}), there exists an inverse trend between $v_\mathrm{nth}$ and peak intensity. The non-thermal velocities here also decay linearly with time but at a faster decay rate of -0.21$\pm$0.02\,km\,s$^{-2}$. The times of increased cosmic ray exposure were excluded from the fit to limit the effects of background noise on the downward trend. In comparison to other investigations into the decay of LT non-thermal broadening in flares \citep[e.g.,][]{stores2021,shen2023_2}, the time scales on which we observe the decay --- either here during the initial signal or that which precedes the brightest \fe emission in Figure \ref{fig:aligned} --- are much shorter, on the order of minutes compared to several tens of minutes. The duration here also results in a rate of decay that is nearly an order of magnitude faster than those measured by \cite{stores2021}. Moreover, the maximum non-thermal velocities observed in this work are also relatively lower than these other works as well, which reach values of 100\,km\,s$^{-1}$ or more. 

In addition to the Gaussian fits, the dynamic binning routine also provides a measure of the LT emission's size and location during the initial signal. The size --- defined by the number of pixels selected, or the height of the red box in Figure \ref{fig:db_frame} --- was found to be roughly constant, with an average size of 3.1\arcsec. The box's center location also deviated slightly over time, traveling northward from 310.1\arcsec to 311.1\arcsec. We note that although our routine bins this region to increase the SNR, the signal of the LT \fe emission in the O\,\textsc{i} detector images (left panels in Figure \ref{fig:db_frame}) is defined by a homogeneous structure that is well approximated by the red boxes. The initial LT emission is therefore well-resolved in our observations, as the 3.1\arcsec\, size of this structure is well over the IRIS SG spatial resolution of 0.166\arcsec. 
Again, this resolution is in contrast to previous investigations into the decay of non-thermal broadenings in flares using Hinode/EIS \citep[e.g.][]{kontar2017,stores2021}, where the resolution, albeit sufficient to locate different regions of excess line widths along the flare loops, could have been affected by averaging over variations of macroscopic plasma dynamics within these individual regions.

The broadening of spectral lines beyond their respective thermal values can have several implications for the plasma dynamics of the emitting region. If we assume the line broadening is due to thermal effects alone, then the large widths in this work would indicate plasma temperatures beyond the expected Log$_{10}\,T_\mathrm{Fe}$[K] = 7.06 thermal formation temperature. For example, the highest value of $v_\mathrm{nth}$ = 66\,km\,s$^{-1}$ is equivalent to a Doppler broadened \fe line formed at Log$_{10}\,T$[K] = 7.41. However, this assumption begs whether it is possible for \fe\,1354.08\,\AA\ to emit at such high temperatures. During flares, rapid heating of tenuous coronal plasma could lead to under-ionized conditions, as the ions in the plasma --- especially highly charged Fe --- will not have time to adjust to equilibrium \cite[i.e., non-equilibrium ionization (NEI)][]{shen2013}. Under-ionization shifts the distribution of charge states, or ionization fractions, amongst different species, giving a bias towards lower plasma temperatures that could result in an underestimation of the actual plasma temperature. Recent work by \cite{shen2023} used three-dimensional MHD simulations to study the effects of NEI on Fe ions during flares, where they found instances of plasma emission dominated by \fe but with true temperatures of more than Log$_{10}\,T$[K]=7.25 --- a difference of over 40\%. While a scenario such as this could help explain some of the large non-thermal broadenings observed in this work, it is unclear whether NEI would be enough to account for such large temperatures. For instance, the 40\% temperature discrepancy in \cite{shen2023} was taken behind a Petschek-type reconnection shock in the current sheet with ambient plasma densities of $5\times 10^8$cm$^{-3}$. As our event is viewed on-disk, emission contributions from the current sheet are likely to be washed out by the denser, underlying arcade plasma. Moreover, the decay in excess line broadening, in either the initial signal or during the time preceding the brightest LT emission, suggests that the plasma is quickly cooling. This scenario would be more consistent with an over-ionized plasma, in which temperatures inferred from dominant Fe species would be overestimated with respect to their equilibrium plasma temperatures (i.e., lower than 11.5\,MK in the case of \fe).
The location of over-ionized plasma in the \cite{shen2023} simulations at the base of the current sheet is also more congruent with the LT emission observed in this work. 

\begin{figure*}
    \centering
    \includegraphics[width = \textwidth]{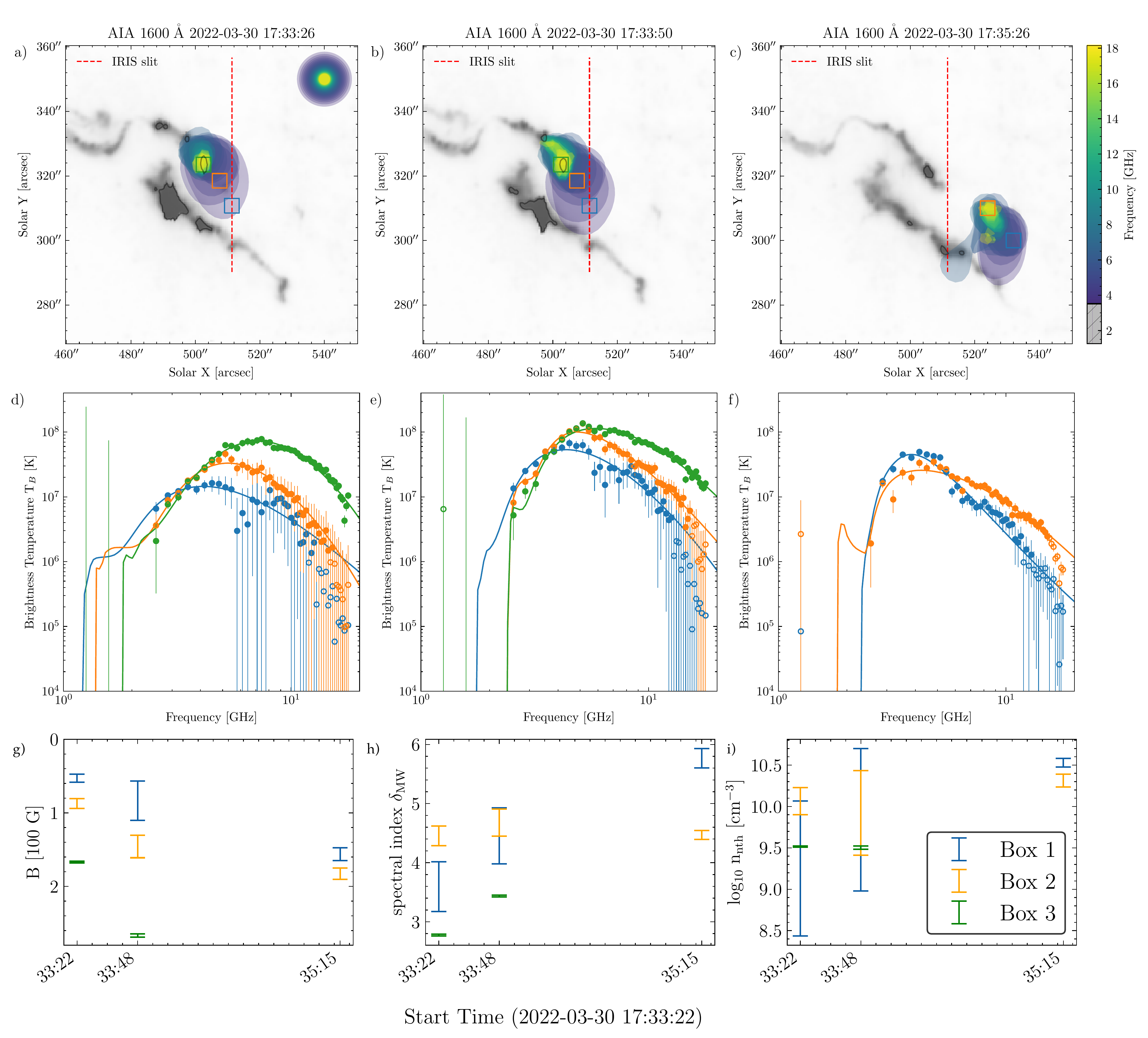}
    \caption{Spatially resolved MW spectra at various locations in the flaring loops at three selected times. Top panels (a)-(c) show inverse greyscale AIA~1600\,\AA\, maps overlaid with 70\% contours from the EOVSA MW map (3.5-18 GHz) at 17:33:22, 17:33:48, and 17:35:15\,UT. Grey contours outline 75\% AIA~1600\,\AA\, flare ribbon intensity. Colored boxes in the top panels correspond to regions selected for spectral analysis, as shown in panels (d-f), with no change in box location between (a) and (b). The filled circles in the upper right corner of panel (a) represent the FWHM size of the restoring beams at the respective frequencies. Middle panels (d)-(f) display the brightness temperature (T$_B$) spectra at these times, color-coded to match the boxes. The colored curves in these spectra represent the best-fit models based on homogeneous gyrosynchrotron emissions from non-thermal electrons following a single power-law distribution, with each curve's color matching that of the corresponding spectra. Open circles in the plots indicate data points excluded for the spectral fitting. Bottom panels (g)-(i) show physical parameter ranges derived from the best-fit homogeneous gyrosynchrotron emission model fits, including magnetic field strength $B$, non-thermal power-law spectral index $\delta_\mathrm{MW}$, and non-thermal electron density $n_\mathrm{nth}$.
    \label{fig:eovsa} }
\end{figure*}

If the LT plasma is assumed to be at the peak formation temperature of \fe, then the excess broadening must instead arise from non-thermal plasma fluctuations, ranging from scales on the order of the emitting ion to large scales associated with bulk mass flows. One popular mechanism proposed for these observations during flares has been the superposition of Doppler-shifted emission from multiple bulk plasma flows. A recent study by \cite{polito2019}, however, used a multi-threaded 1D flare loop model to find that such a superposition results in asymmetric line profiles not commonly observed in \fe during flares. The symmetric line profiles observed here at the resolution of individual IRIS pixels (i.e., Figure \ref{fig:db_frame}) also substantiate this claim. Other mechanisms, such as pressure broadening and optical depth effects, are unlikely to contribute to excess broadening in optically thin UV lines \citep{milligan2011}. The most likely explanation for the excess broadening, therefore, is the superposition of Doppler-shifted emission from unresolved turbulence --- a consensus reached by several studies \citep[e.g.][]{antonucci1984,polito2019,shen2023_2}. Moreover, the decay of $v_\mathrm{nth}$ is also characteristic of MHD turbulence, where the energy cascade from large to small scales results in the dissipation of the turbulent energy contained in the plasma \citep{kontar2017,stores2021}. As we will show in the following sections, it is possible that this energy is being transferred into the acceleration of non-thermal electrons.

\section{Electron Acceleration Signatures} \label{sec:accel}

\subsection{Microwave Imaging Spectroscopy} \label{sec:eovsa}

Full-disk MW images of the flare were captured by EOVSA in 1-18 GHz. Each image used 451 frequency channels distributed across 50 spectral windows with equally-spaced bandwidths of 325 MHz, and was calibrated using standard procedures as described in \cite{chen2020}. Due to a calibration discrepancy in the lowest-frequency spectral windows, images were produced from 45 windows ranging from 3.5-18 GHz. Final images were obtained after subtracting the background between the second and third MW phases at 17:34:50 - 17:35:00\,UT. Image reconstruction was accomplished using the CLEAN algorithm with a frequency-dependent circular beam of size $60\arcsec/\nu_\mathrm{GHz}$.

Preliminary analysis of MW images during the period of initial LT \fe emission found instances where the collection of compact, multi-GHz frequency sources overlapped the IRIS SG slit. Three images were selected for further analysis, corresponding to the start of the initial LT \fe emission, the peak time of the second phase, and the first pulse in the third phase, respectively, and are shown in Figure \ref{fig:eovsa}(a)-(c). The selected times are also indicated in Figure \ref{fig:tmseries}. We see during the first two images that lower-frequency MW sources ($\lessapprox6$\,GHz) are co-spatial with the IRIS slit. Notably, the edge of the MW source in the lowest-frequency spectral window reaches the region of LT \fe emission as it first begins at time 17:33:22 (blue box in panels (a) and (b)). Over time, the MW sources move southwest into and then completely off the slit during the duration of the initial \fe signal.

Beyond the spatiotemporal correlation between the MW and \fe emission, we highlight two points regarding the morphology of the MW sources.
First, the apparent southwestward movement of the MW sources along the arcade is continuous between images (a) and (b) in Figure \ref{fig:eovsa}, which belong to the second pulsation phase, while the transition between (b) and (c) is discontinuous, following a period of decreased MW emission before the start of the third phase. This behavior is consistent with the time-distance stack plots in Figure \ref{fig:stack}, and indicates a spatial variance between the MW pulsations. 
Second, there exists a clear spatial organization between the GHz spectral windows. Previous investigations using EOVSA MW images --- namely those from the 2017 September 10 X8.2 flare \citep[e.g.,][]{gary2018,fleishman2020,yu2020} --- have shown that different GHz spectral windows can correspond to different regions of the flaring arcade, ranging from the lower regions of the flare loop-legs to high in the reconnection current sheet. In our case, the on-disk orientation complicates the interpretation of the MW sources in relation to the flare structure (i.e., loop-legs vs current sheet). Here we see that the higher frequency emission shows a distinct correlation with the intense emission regions of the underlying northern ribbon, shown by the gray 75\% AIA 1600\,\AA\, contours. Moving downwards in frequency, the MW sources reach further southward, seemingly into the LT region of the arcade. A natural interpretation of this organization follows from the standard flare model, where MW sources are tracing out the northern half of a post-flare loop, outlined by the higher frequencies emanating near the ribbon and the lower frequencies originating from higher up in the corona, possibly from the LT or cusp regions.

To assist with this interpretation, we analyzed spatially-resolved MW brightness temperature spectra $T_B$ derived from different locations along the MW sources.
Figure \ref{fig:eovsa} (d)-(f) shows the T$_B$ spectra (closed and open circles) corresponding to the colored boxes in panels (a)-(c), respectively, with box locations chosen to roughly map out our assumed orientation of the MW sources along a post-flare arcade loop. \revise{Considering the beam size may be larger than the boxes drawn, the error of the spectra created for each box was estimated using the image RMS across different frequencies coupled with a systematic error --- 10\% of the maximum brightness temperature (T$_B$) of each frequency --- to accommodate the larger beam sizes at lower GHz.} Each of the eight spectra sampled is consistent with gyrosynchrotron radiation from a population of non-thermal electrons with a power-law energy distribution \citep{dulk1982,chen2020}, exhibiting both positive and negative slopes on the lower and high-end frequencies, respectively. A forward fit to the spectra using the homogeneous gyrosynchrotron emission model described in \cite{chen2020_2} and \cite{fleishman2020} was performed to derive the physical parameters of the non-thermal emission. This routine employs the fast gyrosynchrotron codes outlined in \cite{fleishman2010} and depends on free parameters such as magnetic field strength $B$, non-thermal power-law spectral index $\delta_\mathrm{MW}$, thermal $n_\mathrm{th}$ and non-thermal electron densities $n_\mathrm{nth}$ above $E_\mathrm{min}$=10\,keV. The best fits to these spectra are shown by the solid lines in Figure \ref{fig:eovsa}(d)-(f), where data points having $T_B<1$[MK] or an SNR less than than 1.8 were excluded from the fit (open circles).

Trends in the inferred gyrosynchrotron fit parameters are visualized in the bottom panels of Figure \ref{fig:eovsa}. \revise{Here, the error bars designate the range of accepted values from the fitting routine spanning plus/minus the standard deviation of each model parameter.  Starting with the magnetic field strengths, we find that $B$ is lower in Box 1 and increases sequentially to Box 3, supporting the interpretation that Box 1 represents the LT region and and Boxes 2 and 3 sequentially decrease in height along the flare loop. Values for the power-law spectral index of the non-thermal electron energy distribution $\delta_\mathrm{MW}$ are also consistent with those expected for gyrosynchrotron emission \cite{dulk1982}.} The non-thermal electron density $n_\mathrm{nth}$ is also highest in the \revise{Box 1}, reaching values upwards of $10^{10}$\,cm$^{-3}$. Using these values and in addition to the thermal electron density from the GS model, we found the percentage of non-thermal-to-total electron population to be notable, with \revise{Box 1} ranging from \revise{15-35\%}. This result not only supports the interpretation of the MW sources as gyrosynchrotron emission from non-thermal electrons, but also suggests that the LT is a site of efficient non-thermal electron acceleration during the flare.


\subsection{Hard X-ray Imaging Spectroscopy} \label{sec:obs_stix}

\begin{figure}[!t]
    \centering
    \includegraphics[width = 0.66\textwidth]{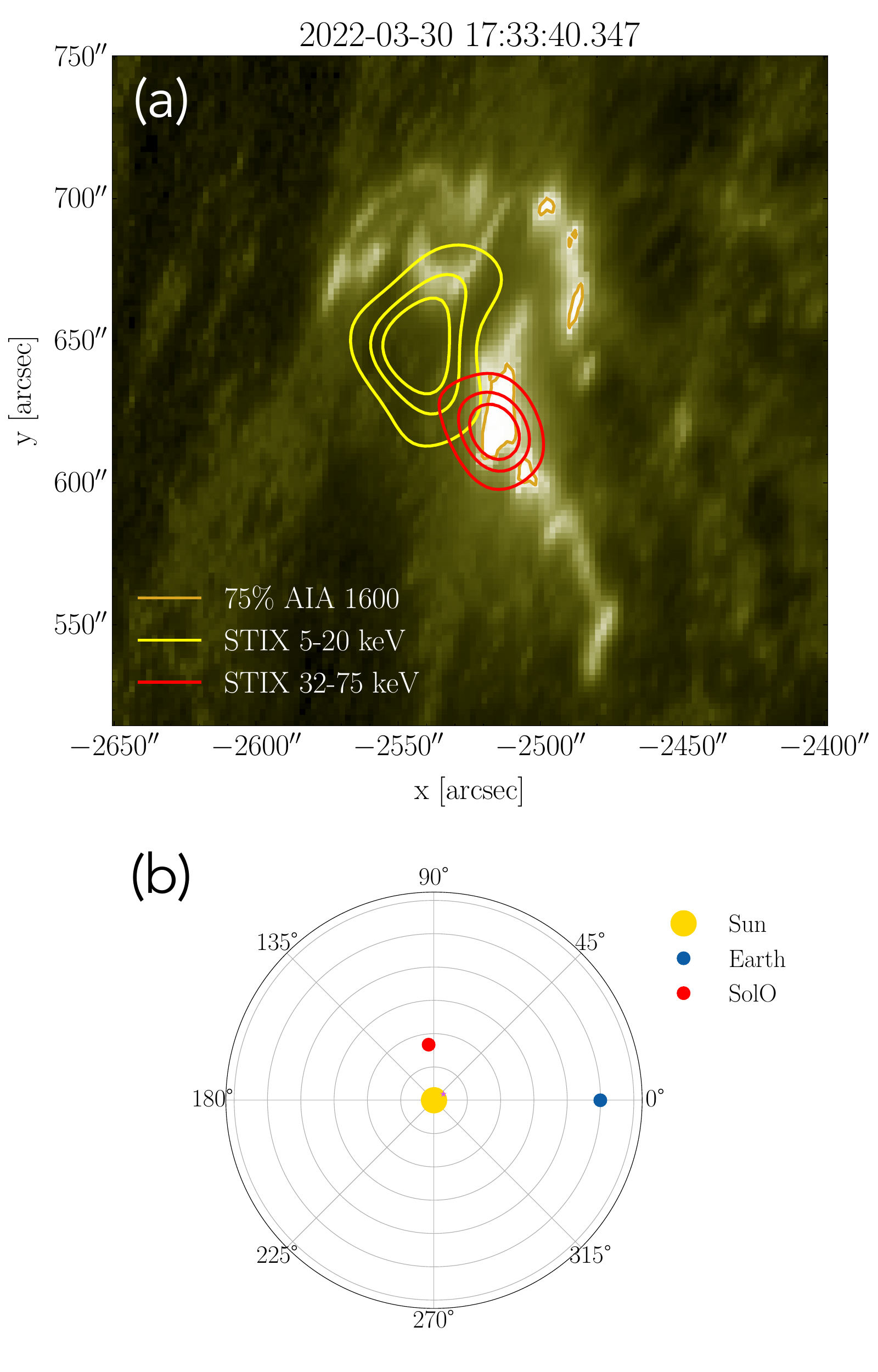}
    \caption{(a) STIX X-ray images integrated from 17:33:14-17:34:06\,UT from point of view of SolO, corresponding to the initial LT \fe emission period. HXR (red) and SXR (yellow) contours at [20,40,60]\% peak emission are overlaid on the corresponding reprojected AIA 1600\,\AA\, image. Orange contours outline 75\% AIA 1600\,\AA\, flare ribbon intensity, and match those shown in Figure \ref{fig:eovsa}. (b) 2022 March 03 Solar Orbiter spacecraft location with respect to Earth in Heliographic Stonyhurst coordinate orientation.
    \label{fig:stix} }
\end{figure}

Additional evidence for electron acceleration comes from HXR data collected by the STIX instrument aboard SolO. A Fourier-imaging spectroscopy instrument, STIX measures X-ray emission in the 4-150\,keV range, and is capable of imaging the Sun with a spatial resolution of $\approx10$\arcsec. Figure \ref{fig:stix}(a) shows SXR (yellow) and HXR (red) contours at [20,40,60]\% peak emission integrated for 48\,s over the first instance of \fe LT emission from 17:33:14-17:34:06\,UT in the Earth's time frame (but from the SolO FOV), with the images reconstructed using the MEM\_GE algorithm \citep{massa2020}. The X-ray contours are overlaid on the corresponding AIA\,1600\AA\ image reprojected to the SolO observing frame, with the contours shifted manually by (-13,
45)\arcsec\, to account for inaccuracies in the STIX aspect system, as was performed in \cite{collier2024}. We find the HXR emission correlates with the intense UV ribbon emission at the southern ribbon, with the northern ribbon showing no significant HXR emission. However, we note the limited dynamic range of STIX ($\sim10:1$) may be overpowering any HXR emission in the northern ribbon. The SolO spacecraft location with respect to Earth at the time of the event is shown in Figure \ref{fig:stix}(b) in Heliographic Stonyhurst coordinates. At the time of this observation, SolO was close to its perihelion with a heliocentric distance of 0.33\,AU.

\begin{figure}
    \centering
    \includegraphics[width = 0.66\textwidth]{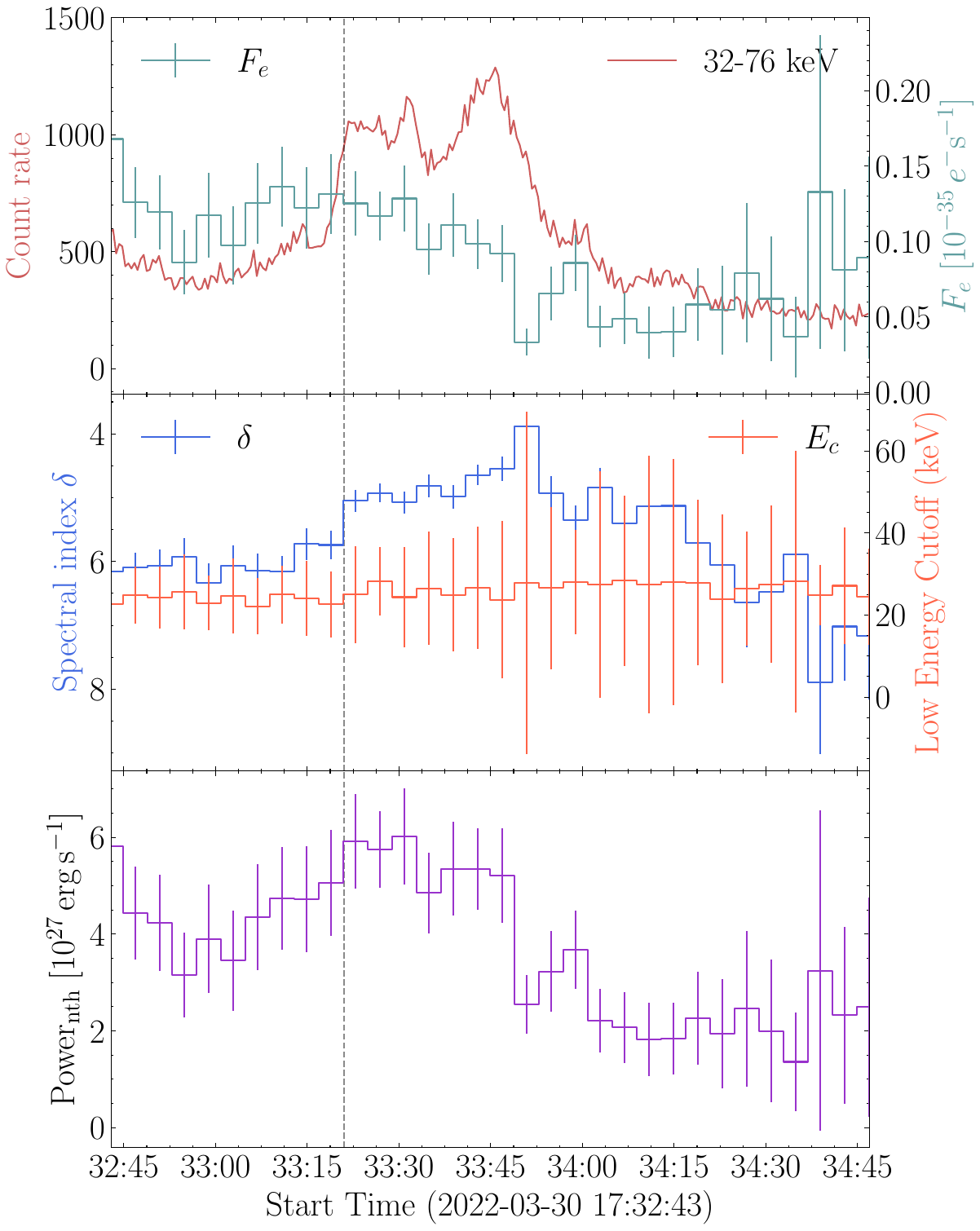}
    \caption{HXR overview of the second phase. Count rate in the STIX 32-75\,keV channel and non-thermal electron flux $F_e$ are plotted in the top panel. The middle panel shows the low energy cutoff ($E_c$) and electron spectral index ($\delta$). Non-thermal electron power is shown in the bottom panel, as calculated by Equation (\ref{eqn:power}) for energies over $E_r$=30\,keV. Values for $F_e$, $E_c$, and $\delta$ follow from best fits to the HXR spectra integrated over 4\,s second intervals. The vertical dashed lines indicate the start time of the initial \fe emission.
    \label{fig:stix_ph2} }   
\end{figure}

The correlation between the HXR and UV ribbon emission is consistent with the model of electron energy transport in flares, where non-thermal electrons accelerated in the corona impact the chromosphere, depositing their energy and producing bremsstrahlung radiation in the form of HXRs \citep{brown1971}. To study the extent of electron deposition in the chromosphere, we performed an analysis of the HXR spectra over the time surrounding the second high-energy phase, also corresponding to the time of the initial LT \fe signal. HXR spectra were fit using the SolarSoft \texttt{OSPEX} routine using a combination of thermal and cold thick-target non-thermal functions (f\_vth and f\_thick2, respectively) at 4\,s intervals over the period matching the second high-energy phase from time 17:32:43-17:34:47\,UT. The resulting non-thermal electron flux, $F_e$ [$e^-$\,s$^{-1}$], low energy cutoff, $E_c$ [keV], and electron spectral index, $\delta$, of the single-power, thick-target component are shown in Figure \ref{fig:stix_ph2}. Here we see a single, steady evolution of the spectral index from soft to hard and back throughout the second phase, which is indicative of a single energy pulsation, or electron acceleration event (in contrast to the many soft-hard-soft pulsations seen in the first phase \citep{collier2024}). Using the best-fit parameters, we can then estimate the power of non-thermal electrons --- those with energy above $E_c$ --- deposited in the chromosphere, $P_\mathrm{nth}$, by 
\begin{equation} \label{eqn:power}
    P_\mathrm{nth}(E \ge E_c) = k_E\; F_e\; E_c\;  \frac{\delta-1}{\delta-2}, 
\end{equation}
where $k_E$ is the energy conversion factor from keV to erg \citep{kontar2019}. 
However, the low energy cutoff measured here is poorly constrained, which adds significant uncertainty to the power estimate. To mitigate this error, we choose a fixed energy reference $E_r$=30\,keV above the mean $\bar{E_c}$=26\,keV and calculate the non-thermal electron power, integrating from $E_r$ to $\infty$. To do this, we use the f\_thick2\_vnorm thick-target function in \texttt{OSPEX}. As a result, Equation \ref{eqn:power} becomes 
\begin{equation} \label{eqn:power}
    P_\mathrm{nth}(E \ge E_r) = \frac{k_E\; A\; E_r^2}{\delta - 2}, 
\end{equation}
where $A$ is the electron flux [$e^-$\,s$^{-1}$\,$\mathrm{keV}^{-1}$] at the reference energy $E_r$ and the other parameters are the same as before. 
While this method underestimates $P_\mathrm{nth}$, it provides a more reliable estimate of the power and its time evolution. The resulting power is shown in the bottom panel of Figure \ref{fig:stix_ph2}, where the trend in $P_\mathrm{nth}$ roughly tracks that of the HXR count rate. The peak of the power also corresponds well with the start time of the initial LT \fe emission (dashed line) --- a connection we explore further in the following section.

\section{Turbulent energy transfer} \label{sec:turb} 

The analysis of non-thermal signatures in this work has revealed excess line-broadening in the \fe line, gyrosynchrotron MW emission, and the deposition of high-energy electrons into the chromosphere --- all of which have been shown to occur concurrently and in close spatial proximity. Together, these signatures are consistent with the acceleration and transport of non-thermal electrons in the LT plasma via a stochastic acceleration mechanism \citep{larosa1993,petrosian2012}. Here we return to the initial non-thermal broadening decay measured in Section \ref{sec:is} to extrapolate on its significance to turbulent dissipation and kinetic energy transfer in light of the associated electron acceleration signatures.

Following the analysis presented in \cite{kontar2017} and \cite{stores2021}, a decrease in excess spectral line broadening can be used to estimate the dissipation of turbulent kinetic energy $K$ in a plasma. Provided a measurement of the non-thermal velocity $v_\mathrm{nth}$, we calculate $K$ by: 
\begin{equation} \label{eq:K}
    K = \frac{3}{2}\;m_i\; v_\mathrm{nth}^2\;n_p\;V,
\end{equation}
where $m_i=1.3\,m_p$ is the mean ion mass given coronal abundances \citep{reames2014}, $n_p$ is the proton number density, and $V$ is the plasma volume of the emitting region.

While $v_\mathrm{nth}$ is directly measured, assumptions must be made to estimate $n_p$ and $V$ in order to calculate the dissipation of kinetic energy. Using the size of the low-GHz MW sources as a proxy for the LT plasma volume at the time of the initial \fe signal (Figure \ref{fig:eovsa}(a)), we estimated the volume of the LT plasma to be $V\approx 3 \times 10^{27}$\,cm$^3$ assuming a cylindrical source with a radius and length of 6\,Mm and 25\,Mm, respectively. We note that the apparent size of the MW sources is biased by the beam size used in their reconstruction (Figure \ref{fig:eovsa}(a)), likely making our rough estimate of the coronal volume the upper limit of the actual MW source size. Nonetheless, the size of the high-frequency MW source from the northern ribbon footprint agrees well with the AIA~1600\,\AA\, ($\approx5$\arcsec, or 3\,Mm) emission there, thereby adding support to our estimation technique. The coronal volume of the SXR source ($\approx 4\times 10^{27}$\,cm$^3$ (Figure \ref{fig:stix}(a)) also agrees with our MW source estimate when using the 50\% contour level. To calculate the proton number density, we assume $n_p = n_e$ following the approach in \cite{kontar2017}, and then use the differential emission measure (DEM) diagnostic to estimate the electron number density $n_e$ in the LT plasma. The details of this analysis are described below.

\subsection{DEM Density Diagnostic} \label{sec:dem}

To understand the evolution of electron number density over time, DEM maps were first created using AIA images in five passbands (94, 131, 193, 211, and 335\AA) over the period 17:29:00-17:53:48\,UT. Due to saturation effects, all 171\,\AA\, images were excluded from the analysis. Images used for a single DEM map were selected at 12\,s intervals within the period, with each channel image selected having the smallest time delta from a given interval. In addition to the AIA image processing described in Section \ref{sec:fe21}, each image was further corrected for degradation and normalized by the exposure time.

DEM calculations in this work were computed using the Python implementation\footnote{Avaliable online at \url{https://github.com/ianan/demreg}} of the regularized inversion method first described in \cite{hannah2012}. 
AIA temperature responses used in the calculations were constructed using a convolution between the effective AIA area taken from the SolarSoft \texttt{aia\_get\_response.pro} routine and atomic data from CHIANTI v10 as described in \cite{delzanna2011}. Because we speculate the initial LT plasma is coronal plasma heated to temperatures above 10\,MK, as opposed to evaporated chromospheric plasma, temperature responses were calculated assuming a coronal abundance \cite{feldman1992}. Each DEM map was internally calculated in emission measure (EM) space and constrained using the EM loci curves of each AIA channel --- a method found to be the most robust following a series of tests of different DEM calculations. Finally, all DEM maps were background subtracted using a DEM map calculated at the pre-flare time 16:45:00\,UT \citep{motorina2020}. Additional details regarding the DEM method, AIA response functions, and abundance assumptions can be found in Appendix \ref{sec:appendix}.

\begin{figure*}
    \centering
    \includegraphics[width = \textwidth]{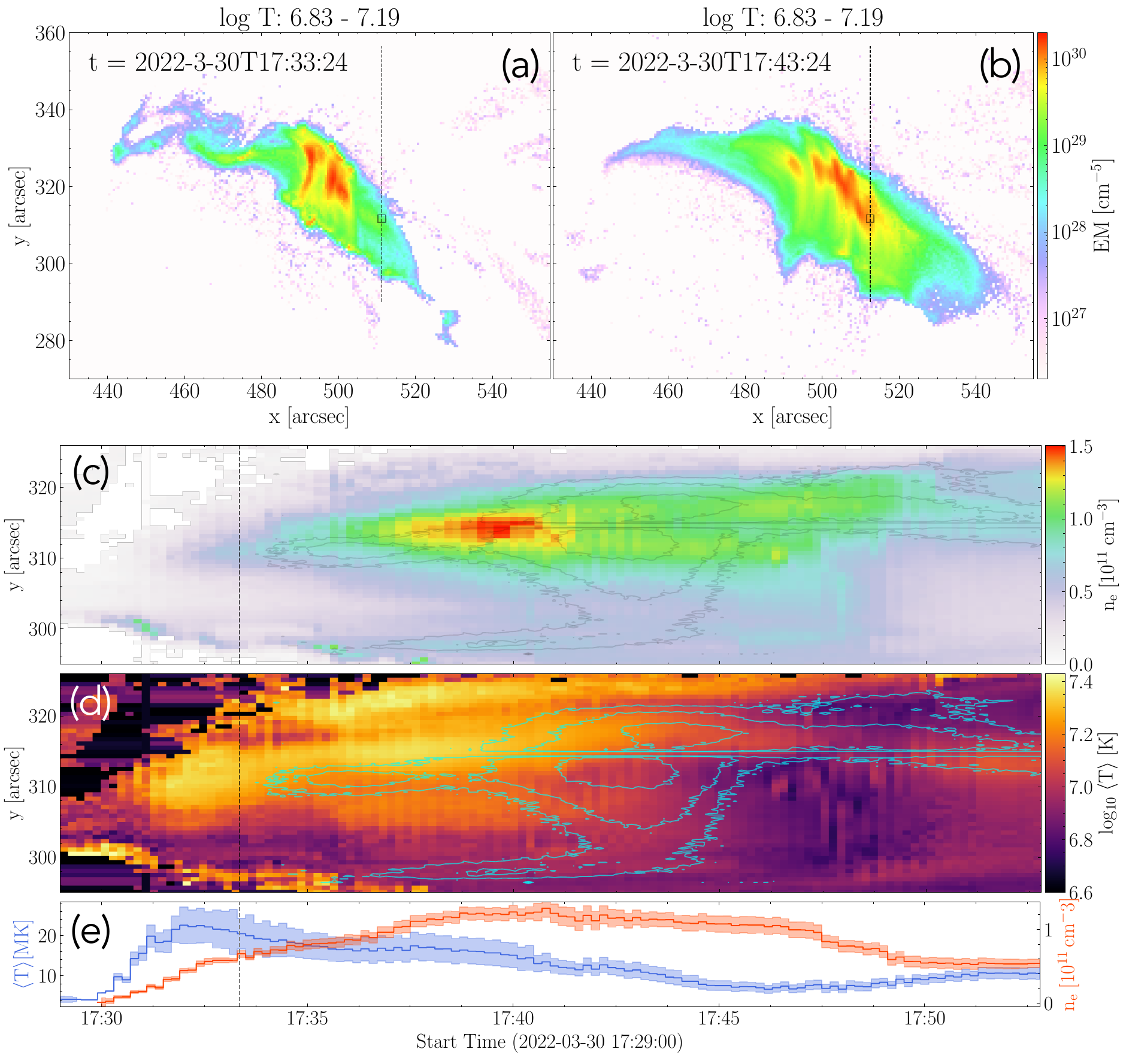}
    \caption{DEM-weighted emission measure (EM) maps in the Log$_{10}$\,[K] 6.83-7.19 range at the start time of the initial signal (a) and peak \fe emission time (b). Dashed lines indicate the IRIS slit position. Black boxes outline the location of \fe LT emission. Time-distance stack plots of the electron number density (c) and DEM-weighted mean temperature  (d) calculated using slices in the EM maps (Equations (\ref{eq:ne}) and (\ref{eq:dem_t}), respectively). Artificial slices made in the EM maps mimic the IRIS slit position over time. Grey and cyan contours are of the Fe xxi intensity levels matching that in Figure \ref{fig:stack}. (e) Time series of the DEM-weighted mean temperature (blue) and electron number density (red), averaged over the area in the black boxes as illustrated in (a) and (b). Vertical dashed lines indicate the start time of the initial signal.
   \label{fig:dem} }
\end{figure*}

Panels (a) and (b) in Figure \ref{fig:dem} show two EM maps calculated from the DEM results at the start time of the initial \fe signal and at the peak time of the \fe emission, respectively, over the range Log$_{10}$\,[K] 6.83-7.19. The dashed lines indicate the IRIS slit position and the black boxes outline the location of the LT \fe emission. At time 17:33:24, there is relatively little EM associated with the initial IRIS \fe emission, which changes as the flare progresses and regions of high EM shift westward over the IRIS slit. To calculate the electron number density $n_e$, we use the following:
\begin{equation} \label{eq:ne}
    n_e = \sqrt{ \frac{EM}{0.83\;\ell} },
\end{equation}
where EM is the total emission measure integrated overall temperature bins in a single DEM map (Log$_{10}$\,[K] 5.7-7.6) and $\ell$ is the LOS column depth of the emitting plasma, assuming a fully-ionized plasma with a relative hydrogen to free electron density ratio of 0.83 \citep{reeves2020,delzanna2021}. Here we take $\ell = 12$\,Mm, following our coronal volume estimate. 

To study the number density evolution across the IRIS slit, we calculate $n_e$ across an artificial slit along the DEM maps matching the SG slit location over time. The resulting time-distance stack plots of $n_e$ are shown in Figure \ref{fig:dem}(c). In addition to $n_e$, we also calculate the DEM-weighted mean temperature $\langle T \rangle$ using the following:
\begin{equation} \label{eq:dem_t}
    \langle T \rangle = \frac{\sum_i DEM(T_i)\; T_i\; \Delta T}{\sum_i DEM(T_i)\; \Delta T} = \frac{\sum_i DEM(T_i)\; T_i\; \Delta T}{EM},
\end{equation}
with the resulting time-distance stack plot shown in Figure \ref{fig:dem}(d). Notably, the density begins to increase at the point $y\approx 311$\arcsec --- consistent with the location of the initial LT \fe emission --- and continues to increase and spread spatially across the SG over time, reaching a peak value of $n_e\approx 1.5 \times 10^{11}$\,cm$^{-3}$ at roughly 17:39:00\,UT. The region of peak density also corresponds to the LT location of the bright EUV knots as seen in Figure \ref{fig:stack}(d)-(f). Preceding the increase in $n_e$ is a rapid increase in the LT plasma temperature, with values $\langle T \rangle$ reaching Log$_{10}$\,[K] $\approx7.4$  at 17:32:00\,UT. The temperature continues to remain elevated above Log$_{10}$\,[K] $\approx7.2$  across the LT region, and gradually decreasing over time.

As a means to directly compare the values of the number density and temperature at the site of the initial \fe non-thermal broadening, we performed a separate DEM calculation on the region corresponding to the black boxes in Figure \ref{fig:dem}(a) and (b). Defined in a 2x2\arcsec\, area centered on the IRIS SG slit over time and $y=311.7$\arcsec, the corresponding AIA pixels were selected to create a 3x3 pixel box, given the 0.6\arcsec\, per pixel resolution of AIA, which was then averaged over. The resulting time series of $n_e$ and $\langle T \rangle$ are shown in Figure \ref{fig:dem}(e). In a consistency check, we find the values of $n_e$ in the range of 0.5-1.5$\times 10^{11}$\,cm$^{-3}$ are in agreement with the values of thermal electron population density derived from EOVSA MW spectra (Figure \ref{fig:eovsa}). Additionally, using the EM$_\mathrm{SXR}$ from the f\_vth of the thermal STIX source and our coronal volume estimate, we find the number density inferred from the SXRs are in the range $n_e = \sqrt{EM_\mathrm{SXR}/V} = [0.3-0.5]\times 10^{11}$\,cm$^{-3}$ during the period of the initial \fe signal, which also agrees with our DEM estimate. We further note that the calculated densities on the order of $10^{10-11}$\,cm$^{-3}$ at the time of the initial \fe emission suggest that any effects due to NEI on the \fe line formation would likely be small. 


\subsection{Kinetic Energy Dissipation}

\begin{figure}
    \centering
    \includegraphics[width = 0.8\columnwidth]{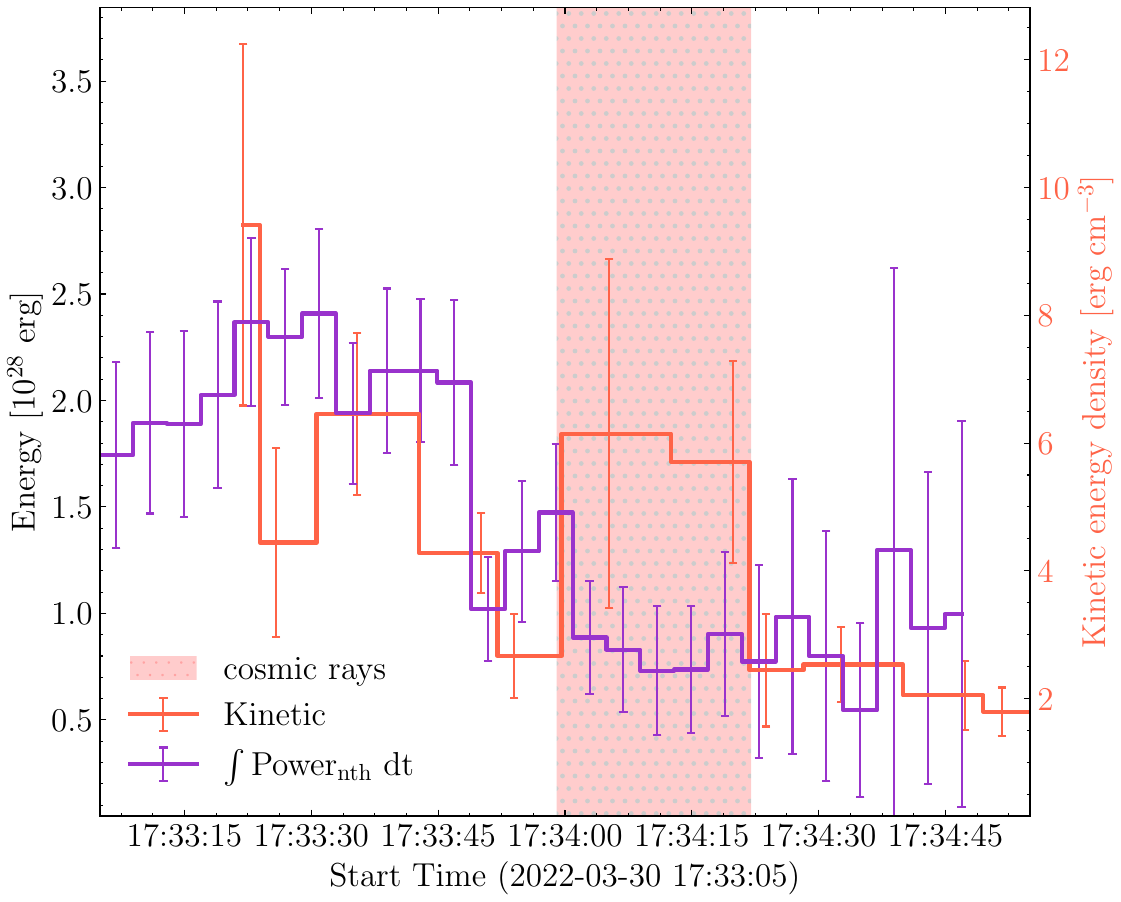}
    \caption{Time series of the turbulent kinetic energy inferred from IRIS \fe non-thermal broadening during the initial signal (orange) and the corresponding time-integrated non-thermal power from STIX HXR emission during the second energy phase (purple). A kinetic energy density scale is also provided for the turbulent kinetic energy on the right axis.
    \label{fig:energy} }
\end{figure}

Following the estimate of the electron number density evolution in the LT plasma above, the dissipation of turbulent kinetic energy in the initial \fe signal is calculated using Equation (\ref{eq:K}). Using the time series of the non-thermal velocity $v_\mathrm{nth}$ inferred from the dynamic binning routine (Figure \ref{fig:fexxi_tmseries}) and the associated electron number density $n_e$ from the DEM analysis, the resulting time series of $K$ is shown in Figure \ref{fig:energy} (orange). The peak kinetic energy is found to be $2.8\pm0.8\times10^{28}$\,erg, which then decreases to $0.6\pm0.1\times10^{28}$\,erg over the next 100\,s. As $n_e$ is increasing over this period, from $0.7\times10^{11}$ to $0.9\times10^{11}\;\mathrm{cm}^{-3}$, the decay in $K$ is slower than the decay in $v_\mathrm{nth}$ inferred in Section \ref{sec:is}. Furthermore, as the total kinetic energy calculation relies on an estimate of the plasma volume, values of kinetic energy density ($K/V$) are also provided in Figure \ref{fig:energy}, read off the right axis, with energy densities roughly on the order of 2-9\,erg\,cm$^{-3}$. This reduction allows for a more direct comparison to the values derived in \cite{stores2021}, who found kinetic energy densities nearly an order of magnitude lower using Fe\,\textsc{xxiv}\,255\,\AA\, but for an off-limb X1.2-class flare. However, this discrepancy is likely due to the lower values of $n_e$ calculated in their work ($\approx10^{9}$ cm$^{-3}$ in regions of high $K$), as similar magnitudes of $v_\mathrm{nth}$ were found in both studies. 

To put the kinetic energy dissipation in context, we compare it to the power imparted on the chromosphere by non-thermal electron deposition during the second energy phase (bottom panel Figure \ref{fig:stix_ph2}). Time integrating $P_\mathrm{nth}$ at each 4\,s interval, the resulting energy time series is shown in Figure \ref{fig:energy} (purple). The non-thermal energy peaks at $2.4\pm0.4\times10^{28}$\,erg, which is consistent with the peak of the turbulent kinetic energy. While the cadence of the IRIS observation is higher than that of the \texttt{OSPEX} time-integrated HXR fitting parameters, there is also a remarkable agreement between the evolution of the two energy estimates starting at the beginning of the initial \fe signal and lasting until the end of the second phase. Despite this agreement, we note that because the reference energy $E_r$ is used to calculate our non-thermal electron power $P_\mathrm{nth}$, which is larger than the inferred high-energy cutoffs $E_c$ during this period (Figure \ref{fig:stix_ph2}), the values of $P_\mathrm{nth}$, and subsequently the integrated non-thermal election energy, are likely underestimated to a slight degree.

\begin{figure}
    \centering
    \includegraphics[width = 0.8\columnwidth]{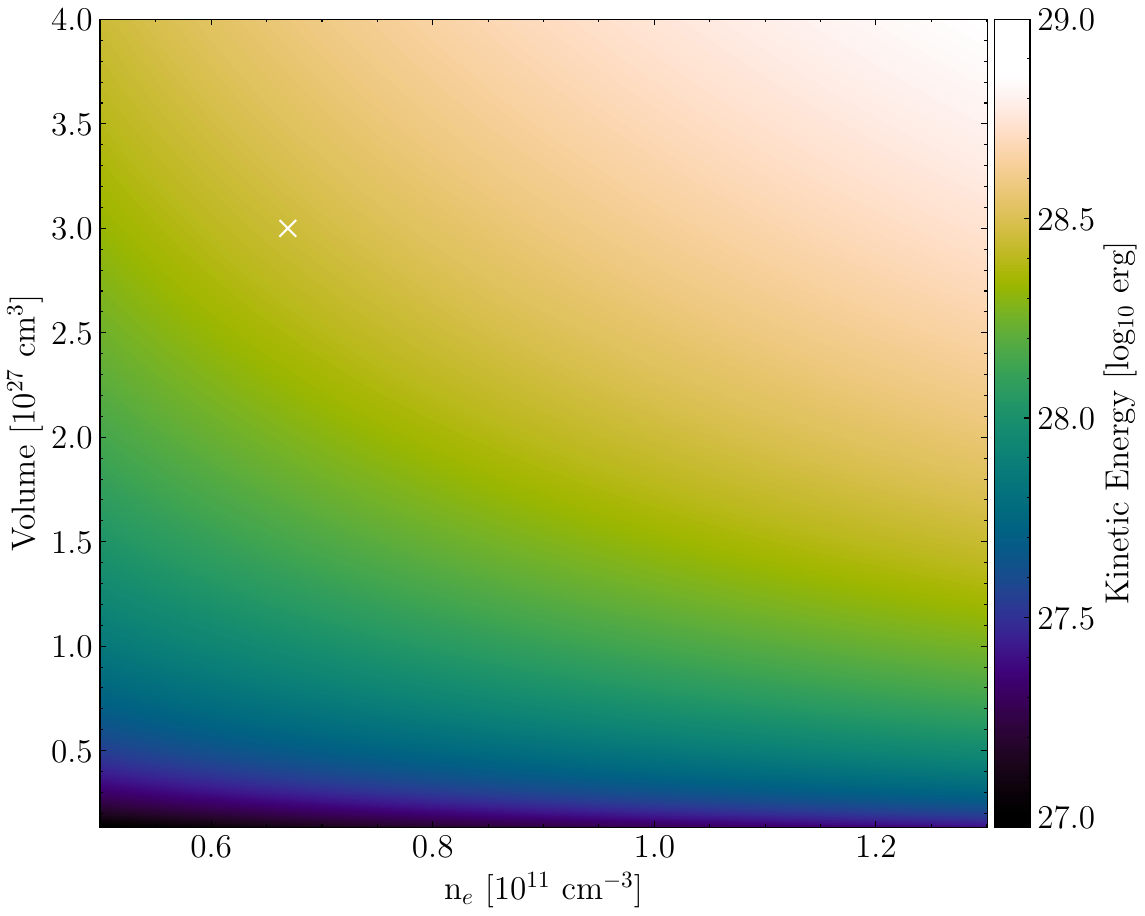}
    \caption{Parameter space exploration of kinetic energy using the peak non-thermal velocity of the IRIS \fe line during the initial signal $v_\mathrm{nth}$=65.7\,km\,s$^{-1}$ for a range of hypothetical number densities $n_e$ and coronal volumes calculated using Equation (\ref{eq:K}). The white marker indicates the number density at the time of peak $v_\mathrm{nth}$, Log$_{10}\,n_e$ [cm$^{-3}$] = 10.83, and coronal volume estimate used in this work, $V = 3\times10^{27}$ cm$^3$.
    \label{fig:volumes} }
\end{figure}

Assuming our coronal volume estimate is accurate, the agreement between the two energy curves in Figure \ref{fig:energy} suggests the dissipation of turbulent kinetic energy in the LT plasma is enough to drive the acceleration of non-thermal electrons into the chromosphere. Not only does this connection hint at a causal relationship, but it also suggests that energy transfer from turbulence to electron acceleration is quite efficient, given the virtual lack of any time delay between the driver (turbulence) and the response (non-thermal electron deposition). This efficiency is supported by numerical experiments of stochastic acceleration in turbulent flare plasmas \citep{miller1996,stackhouse2018}, which have demonstrated sub-second timescales for particle acceleration. However, our coronal volume estimate assumes the kinetic energy derived from the \fe non-thermal broadening is contained within a volume representative of the MW and SXR sources. As we showed in Section \ref{sec:is}, the spatial extent of the initial \fe signal along the IRIS slit is $\approx 3.1$\arcsec --- a fraction of the overall extent of the MW emission. While the smaller size of the \fe emission implies that the turbulent kinetic energy may be overestimated, the IRIS slit may only be sampling a small portion of the turbulent LT plasma responsible for the electron acceleration. 
Nonetheless, when combined with the possibility of an underestimated range of $P_\mathrm{nth}$, we cannot neglect the possibility that the turbulent kinetic energy inferred by IRIS is less than the energy of chromospheric electron deposition measured by STIX, suggesting turbulence is not the sole acceleration mechanism at work. The dependency of the turbulent kinetic energy $K$ on the coronal volume $V$ is put into context in Figure \ref{fig:volumes}, where the observed peak \fe non-thermal velocity value is used to calculate possible values of $K$ using an estimated range of coronal volumes, as well as electron number densities $n_e$ found from our DEM analysis (Figure \ref{fig:dem}(e)). The volume range is constructed such that the lower bound is a cylindrical volume calculated using a diameter of 3.1$\arcsec$ from the IRIS observation and the upper bound is the volume of the coronal SXR source as measured by STIX. 

In addition to the non-thermal energy derived from HXR emission, a comparison can be made between the turbulent kinetic energy density and the non-thermal energy density inferred from the EOVSA MW spectra. As detailed in \cite{fleishman2020}, the non-thermal energy density $w_\mathrm{nth}$ is calculated using the following:
\begin{equation}
    w_\mathrm{nth} = k_E\,n_\mathrm{nth}\,E_\mathrm{min}\; \frac{\delta_\mathrm{MW}-1}{\delta_\mathrm{MW}-2},
\end{equation}
given the best-fit parameters from the MW spectra shown in Figure \ref{fig:eovsa} and a minimum energy cutoff of $E_\mathrm{min}$=10\,keV used in the GSFIT routine. The resulting non-thermal energy density was found to be highest in the LT region, increasing from 4.7$\times10^1$ to 2.4$\times10^2$\,erg\,cm$^{-3}$ between times 17:33:22-17:33:48 (Figure \ref{fig:eovsa}). This increase in $w_\mathrm{nth}$ occurs concurrently with the decay in kinetic energy, fitting the narrative of non-thermal electron activity driven by the dissipation of turbulence. However, the MW non-thermal energy density attains values nearly two orders of magnitude higher than the kinetic energy densities inferred from the \fe non-thermal broadening. 

This second order of magnitude discrepancy between the non-thermal and kinetic energy densities was also observed in \citep{fleishman2020}, who  
attributed the difference to the emission regions from which the MW and EUV --- responsible for the two energy estimates --- emanate. In their work, which focused on the September 10th, 2017 event, MWs were observed to stem from the cusp region near the LTs and large values of non-thermal broadening from the Fe\,\textsc{xxiv}\,255\,\AA\, EIS line ($\approx100$\,km\,s$^{-1}$) was primarily confined to the overlying plasma sheet (see also \cite{warren2018,polito2018}), suggesting that the cusp region contained a higher degree of non-thermal electron activity. Although the on-disk nature of our observation hinders a direct comparison between flare geometries, the location of the MW sources in Figure \ref{fig:eovsa} in combination with the HXR footpoint emission in Figure \ref{fig:stix} creates a coherent picture of a loop-like structure, where the MW sources trace out one half of the loop from the corona (low-GHz) to the northern ribbon footpoint (high-GHz) and the HXR emission highlights the location of the southern ribbon footprint (This picture is also consistent with a magnetic mirroring effect, where accelerated electrons reflect off high-$B$ regions in the northern ribbon and precipitate into the southern ribbon, possibly explaining the opposing asymmetry seen in the MW and HXR sources).

\begin{figure*}
    \centering
    \includegraphics[width = \textwidth]{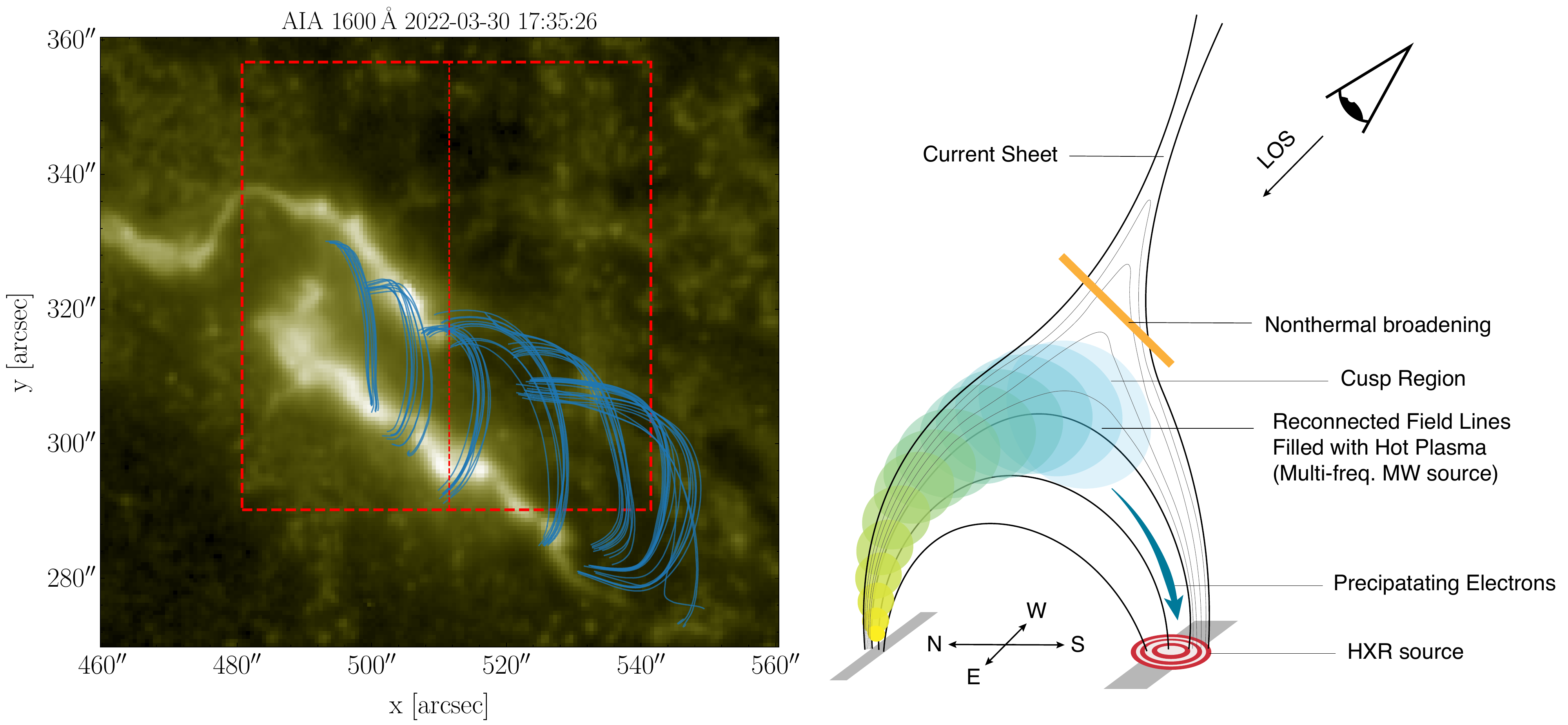}
    \caption{\revise{Left: Magnetic field extrapolation (blue lines) overplotted on AIA\,1600\,\AA\,image at 17:35:26, along with IRIS FOV and SG slit (red-dashed lines). Right:} Schematic diagram showing the interpreted locations of the different observations along the line-of-sight (LOS) as they pertain to the standard flare model. Microwave (MW) sources are confined to the closed, reconnected post-flare loops, which trace out the northern half of the loop-like structure. High-frequency sources (yellow) stem from the northern ribbon, while the low-frequency sources (blue) emanate from the LT/Cusp region. The hard X-ray (HXR) source marks the location of the southern footprint, arising from the precipitation of accelerated electrons. Lying above the MWs in the LT/Cusp region, near the location where the current sheet meets the arcade, is the \fe non-thermal broadening detected along the IRIS SG slit. 
    \label{fig:schematic} }
\end{figure*}

If we adopt the reasoning presented in \cite{fleishman2020}, then it follows the non-thermal broadening in the \fe should lie above the MW sources, either higher up in the LT/cusp region or reaching into the current sheet. We note that this location of \fe non-thermal broadening also agrees with simulations in \cite{shen2023_2}, who found the source of large non-thermal broadening in the IRIS \fe line to be the highly turbulent plasma located along the current sheet and LT/Cusp region. The location of the SG slit shown in Figure \ref{fig:eovsa}, slightly to the right of the MW sources, further supports the notion of vertical separation between the non-thermal signatures measured by IRIS and EOVSA when considering the projection of the arcade on the disk and the LOS of these observations. A schematic of this interpretation is illustrated in Figure \ref{fig:schematic}. \revise{The configuration of the drawn flare loop orientation is further justified by that produced from a magnetic field extrapolation (left panel) performed using the GX simulator modeling package \citep{nita2023}, which employed a nonlinear force-free field reconstruction of the magnetic field from the preflare SDO/HMI magnetogram at 17:23:36\,UT.}

\revise{
Altogether, these findings indicate the co-spatiotemporal presence of non-thermal signatures in the LT and cusp regions. We interpret the decay in non-thermal velocity as the dissipation of turbulent energy in these regions, which can provide the required energy necessary to accelerate electrons into the lower solar atmosphere. Nonetheless, because the SG slit is positioned over the flare arcade at a single location, we can only infer that stochastic acceleration occurs at the time and place in which we observe the initial decay in \fe broadening and the associated MW and HXR signals. In other words, our analysis only pertains to the second high-energy phase (see Figure \ref{fig:tmseries}), where we see the compact MW source move into the SG slit and not the entire flare. Moreover, the decay in non-thermal velocity suggests that the turbulence in the LT/Cusp region is dissipating and no longer being generated --- a notion potentially at odds with investigations that see persistent reconnection outflows generating turbulence for sustained periods \citep{shibata2023,wang2023_2}. While it is likely that reconnection is ongoing following the generation of the \fe broadening, as evidenced by the third high-energy phase, the movement of the MW and HXR sources westward along the flare arcade suggests that the mechanisms responsible for their formation \citep[e.g. reconnection outflows,][]{yu2020} are discrete and localized along the current sheet. The generation of turbulence observed in the \fe line under this assumption would not be ongoing at the location of the IRIS slit and would therefore be allowed to decay freely. An analysis of such extended outflows, albeit beyond the scope of the present work, would be a promising focus for future research.
}

\section{Summary and Conclusion} \label{sec:disc}

The X1.3 class flare on March 30th, 2022 analyzed in our study exhibited non-thermal signatures across measurements taken by IRIS, EOVSA, and STIX. Our findings highlight significant excess broadening of the \fe\,1354.08\,\AA\, spectral line emanating from the flare arcade's LT region, reaching values upwards of 65\,km\,s$^{-1}$. These non-thermal velocities were also observed to decay over time at two different periods, beginning at the onset of LT emission and before peak line emission. We found the evolution of the latter period to coincide with the appearance of bright EUV knots in AIA 94, 131, and 193\,\AA\, channels, offering a new spectroscopic insight into this phenomenon.

Our observations of the LT \fe emission present a unique perspective on the dynamics of solar flares, with the high spatial resolution of IRIS and a fast cadence of $\approx9$\,s, allowing for the direct measurement of non-thermal broadening in a well-resolved LT structure. The rapid decay of non-thermal broadening in the LT plasma, on the order of minutes, is significantly quicker than the tens of minutes reported in earlier studies. 
Moreover, confirming the LT emission from a well-resolved structure challenges earlier works with lower resolution spectrometers and lower cadence where flare emissions might have been interpreted as a superposition of flows from different macroscopic locations along the flare arcade and LOS. 
This highlights the importance of spatial resolution and cadence in understanding solar flare dynamics.

Excess broadening in the \fe line was interpreted as the presence of turbulent plasma in the LT region, with its decay indicative of energy dissipation via a turbulent cascade. Following methodologies from \cite{kontar2017}, we estimated the LT plasma's peak turbulent kinetic energy to be approximately $2.8\pm0.8\times10^{28}$\,erg. This energy correlated well with the time-integrated non-thermal electron power $P_\mathrm{nth}$ deposited in the chromosphere, as inferred from STIX HXR measurements, which peaked at $2.4\pm0.4\times10^{28}$\,erg. As the temporal decay in turbulent kinetic energy also matched that of the integrated electron power, the agreement between the two energies suggests a relationship between the dissipation of turbulent energy and the acceleration of non-thermal electrons, thereby supporting stochastic acceleration theories. We note, however, that this interpretation is contingent on two factors: first, the assumption of a coronal volume estimate, which may overestimate the turbulent kinetic energy as observed by IRIS; and second, the values of $P_\mathrm{nth}$, which are potentially underestimated in our calculations. The

The connection between the turbulent plasma and non-thermal electron populations was further supported by the simultaneous presence of gyrosynchrotron MW emission from EOVSA, where the MW sources were found to be co-spatial and co-temporal with the initial LT \fe emission. Analysis of MW spectra at different times and locations along the flare arcade indicated a significant non-thermal electron fraction in the LT plasma, with a ratio of non-thermal to thermal electrons \revise{ranging upwards of 35\%}.  Although this high percentage indicates an efficient electron acceleration mechanism at work and also places plasma turbulence in the vicinity of these non-thermal electrons at the flare LT, the non-thermal energy density inferred from the MW sources was found to be substantially higher than the turbulent kinetic energy density inferred from the \fe broadening. This discrepancy, first seen in \cite{fleishman2020}, supports the idea of a vertical variation in the non-thermal energy density across the flare arcade, where the formation region of \fe belongs to a region of lesser non-thermal energy density, possibly higher in the solar corona than the underlying MW emission, as summarised in our cartoon in Fig.~\ref{fig:schematic}.

Together, these findings contribute to a deeper understanding of non-thermal processes in the flare LT region, setting a foundation for future investigations and modeling efforts. The forthcoming MUSE mission, with its advanced EUV spectroscopic capabilities and 35 slits \citep{cheung2022}, promises to enhance our ability to study these processes in greater detail.


\begin{acknowledgments}

W.A. is supported by NASA contract NNG09FA40C (IRIS). VP acknowledges support from  NASA Heliophysics System Observatory Connect  Grant \#80NSSC20K1283,\revise{NASA ROSES Heliophysics Guest Investigator program Grant \#80NSSC20K0716,} and from NASA under contract NNG09FA40C (IRIS).  S.Y. is supported by NASA grant 80NSSC20K1318 and NSF grant AST-2108853 awarded to NJIT. H.C. is supported by the Swiss National Science Foundation Grant 200021L\_189180 for STIX. L.A.H is supported by an ESA Research Fellowship. 
IRIS is a NASA small explorer mission developed and operated by LMSAL with mission operations executed at NASA Ames Research Center and major contributions to downlink communications funded by ESA and the Norwegian Space Center. 
EOVSA operation is supported by NSF grant AST1910354 and AGS-2130832 to NJIT. 
Solar Orbiter is a mission of international cooperation between ESA and NASA, operated by ESA. 
The STIX instrument is an international collaboration between Switzerland, Poland, France, Czech Republic, Germany, Austria, Ireland, and Italy. 
AIA and data are provided courtesy of NASA/SDO and the AIA science team. This work has benefited
from the use of NASA’s Astrophysics Data System. 
CHIANTI is a
collaborative project involving researchers at the universities of
Cambridge (UK), George Mason, and Michigan (USA). 
This research used version 5.1.1 of the SunPy open-source software package \citep{sunpy_community2020}.
\revise{The authors thank the referee for suggestions that improved the manuscript.}
\end{acknowledgments}

\appendix

\section{DEM Methods} \label{sec:appendix}

\begin{figure}
    \centering
    \includegraphics[width = 0.66\columnwidth]{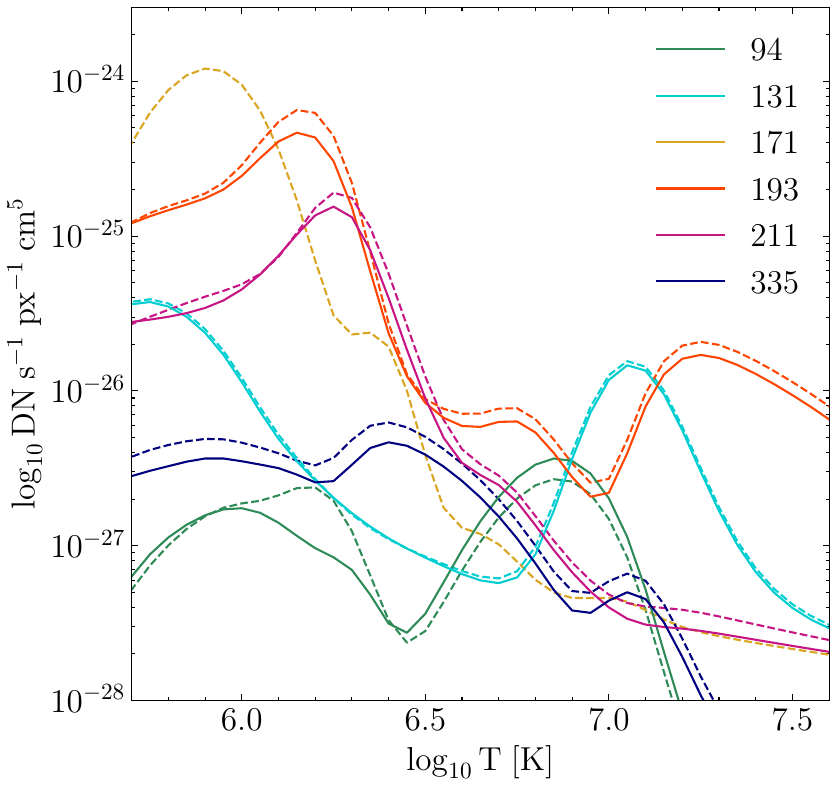}
    \caption{SDO AIA temperature response functions calculated using the SolarSoft \texttt{aia\_get\_response.pro} routine (dashed) and using the method described in \cite{delzanna2011}.
    \label{fig:aia_resp} }
\end{figure}

Figure \ref{fig:aia_resp} shows the AIA temperature response functions used for the DEM calculations in Section \ref{sec:dem} (solid). These functions were calculated using the convolution method described in \cite{delzanna2011} with ion abundances calculated using the CHIANTI v10 atomic database and assuming coronal abundances \cite{feldman1992}. We note that the response function for the 171\,\AA\, channel was not calculated, as the channel was excluded from the DEM calculations due to saturation effects. Temperature response functions calculated using the SolarSoft \texttt{aia\_get\_response.pro} routine are also shown in Figure \ref{fig:aia_resp} (dashed) for comparison.

\begin{figure}
    \centering
    \includegraphics[width = 0.66\columnwidth]{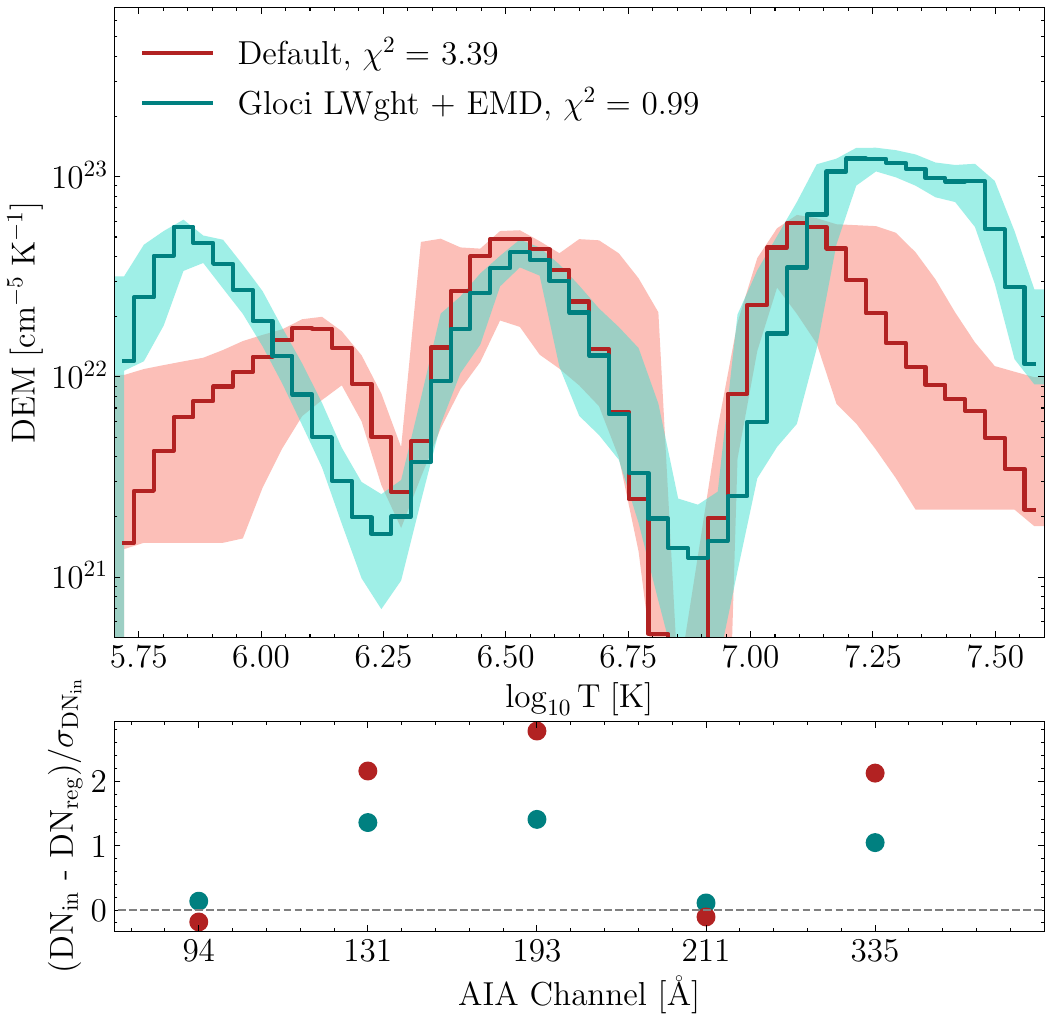}
    \caption{Top Panel: DEM calculation results for the pixels averaged within the box in Figure \ref{fig:dem}(a) at time 17:33:24, assuming a coronal abundance. The results of the regularized inversion method are shown in red and the EM loci constrained results are shown in blue. Bottom Panel: Standardized residuals between the predicted emission from the DEM results and the input AIA data for each method.
    \label{fig:dem_coronal} }
\end{figure}

As AIA 171\,\AA\, images were not included in our analysis, we are consequently neglecting any emission contributions from the Fe IX  171\,\AA\, line (peak Log$_{10}\,T$ [K] =5.9). To ensure the validity of the DEM calculations in the absence of the 171\,\AA\, channel, consistency checks were performed by comparing the predicted emission from the DEM results to the input AIA data. An example of this check is shown in Figure \ref{fig:dem_coronal} for the pixels averaged within the box in Figure \ref{fig:dem}(a) at time 17:33:24 for two inversion methods: the regularized inversion method described in \cite{hannah2012} and the same method constrained by the EM loci curves of the five AIA channels, calculated internally in EM space. While both methods were able to reproduce the observed AIA emission well in the absence of the 171\,\AA\, channel, the EM loci-constrained method was found to be the most robust with a reduced $\chi^2$ of 0.99. 

\begin{figure}
    \centering
    \includegraphics[width = 0.66\columnwidth]{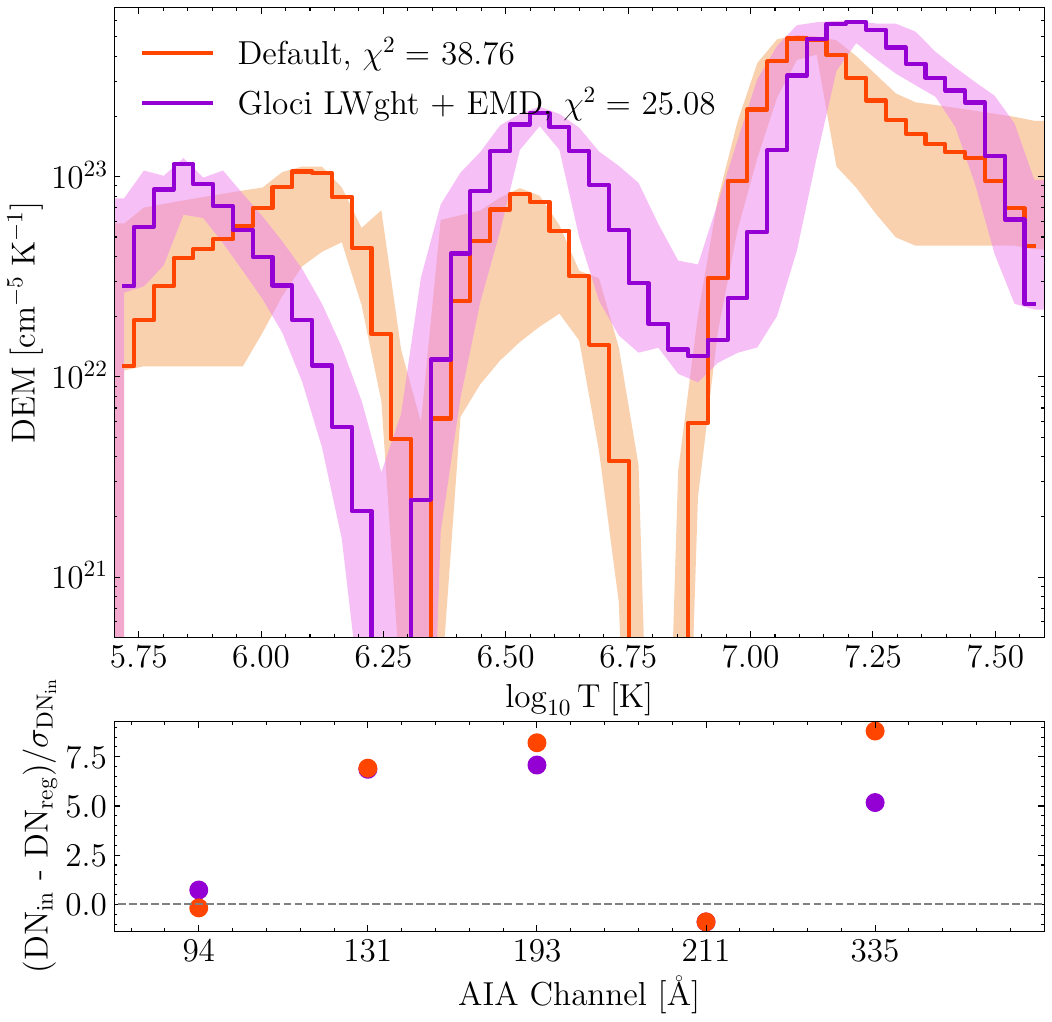}
    \caption{DEM calculation results and standardized residuals assuming a photospheric abundance, plotted as in Figure \ref{fig:dem_coronal}.
    \label{fig:dem_photo} }
\end{figure}

An additional consistency check was done to test our assumption of coronal abundances, given that the ablation of chromospheric material into the corona from evaporation is likely to occur during flares. Performing the same DEM calculations above, but using AIA temperature response functions recalculated with a photospheric abundance \cite{asplund2009}, we found both methods performed considerably worse than those using a coronal abundance (Figure \ref{fig:dem_photo}). However, while this test agrees with our initial assumption for the use of DEMs in this paper, we acknowledge that a more comprehensive study of the DEM calculations using different abundances is beyond the scope of the current work.

\bibliography{ref}{}
\bibliographystyle{aasjournal}

\end{document}